\documentclass[12pt,preprint]{aastex}
\usepackage{graphicx}
\usepackage{amsmath,amssymb}
\usepackage{natbib}
\usepackage{multirow}

\shorttitle{Advective vs Diffusive Solar Convection Zones}
\shortauthors{Yeates et al.}

\newcommand{\pdiff}[2]{\frac{\partial #1}{\partial #2}}

\newcommand{\pdlin}[2]{\partial #1/\partial #2}
\renewcommand{\eqref}[1]{(\ref{eqn:#1})}
\newcommand{\mpsec}{\,\textrm{m}\,\textrm{s}^{-1}}
\newcommand{\cmsqpsec}{\,\textrm{cm}^2\,\textrm{s}^{-1}}
\newcommand{\permet}{\,\textrm{m}^{-1}}
\newcommand{\met}{\,\textrm{m}}
\newcommand{\erf}{\,\textrm{erf}}
\newcommand{\bpol}{\mathbf{B}_{\textrm{p}}}
\newcommand{\mx}{\,\textrm{Mx}}

\begin{document}

\title{Exploring the Physical Basis of Solar Cycle Predictions:\\
 Flux Transport Dynamics and
Persistence of Memory in Advection versus Diffusion Dominated Solar Convection Zones}

\author{Anthony~R.~Yeates}
\affil{School of Mathematics and Statistics, University of St Andrews, St Andrews, KY16 9SS, UK}
\email{anthony@mcs.st-and.ac.uk}
\author{Dibyendu Nandy}
\affil{Department of Physics, Montana State University, Bozeman, MT 59717, USA}
\email{nandi@mithra.physics.montana.edu}
\and
\author{Duncan~H.~Mackay}
\affil{School of Mathematics and Statistics, University of St Andrews, St Andrews, KY16 9SS, UK}
\email{duncan@mcs.st-and.ac.uk}


\begin{abstract}

The predictability, or lack thereof, of the solar cycle is governed by numerous
separate physical processes that act in unison in the interior of the Sun. Magnetic
flux transport and the finite time delay it introduces, specifically in the so-called
Babcock-Leighton models of the solar cycle with spatially segregated source regions for
the $\alpha$ and $\Omega$-effects, play a crucial rule in this predictability. Through
dynamo simulations with such a model, we study the physical basis of solar cycle
predictions by examining two contrasting regimes, one dominated by diffusive magnetic
flux transport in the solar convection zone, the other dominated by advective flux
transport by meridional circulation. Our analysis shows that diffusion plays an
important role in flux transport, even when the solar cycle period is governed by the
meridional flow speed. We further examine the persistence of memory of past cycles in
the advection and diffusion dominated regimes through stochastically forced dynamo
simulations. We find that in the advection-dominated regime, this memory persists for
up to three cycles, whereas in the diffusion-dominated regime, this memory persists for
mainly one cycle. This indicates that solar cycle predictions based on these two
different regimes would have to rely on fundamentally different inputs -- which may be
the cause of conflicting predictions. Our simulations also show that the observed solar
cycle amplitude-period relationship arises more naturally in the diffusion dominated
regime, thereby supporting those dynamo models in which diffusive flux transport plays
a dominant role in the solar convection zone.
\end{abstract}


\section{Introduction}

Direct observations of sunspot numbers over 400 years, as well as proxy data for much
longer timescales \citep{beer2000}, show that both the amplitude and the duration of
the solar magnetic cycle vary from one cycle to the next. The importance of this
phenomenon lies in the contribution of varying levels of solar activity to long-term
climate change, and to short-term space weather \citep{nandy2007}. While there is now a
concensus that the Sun's magnetic field is generated by a hydromagnetic dynamo
\citep{ossendrijver2003,charbonneau2005}, the origin of fluctuations in the basic cycle
is yet to be conclusively determined. Several different mechanisms have been proposed,
including nonlinear effects \citep{tobias1997, beer1998, knobloch1998, kueker1999,
wilmotsmith2005}, stochastic forcing
\citep{choudhuri1992,hoyng1993,ossendrijver1996,charbonneau2000,mininni2002}, and
time-delay dynamics \citep{yoshimura1978,durney2000,charbonneau2005b,wilmotsmith2006}.
A coupled, equally important, but ill-understood issue is how the memory of these
fluctuations, whatever may be its origin, carries over from one cycle to another
mediated via flux transport processes within the solar convection zone (SCZ). A
unified understanding of all these disparate processes lays the physical foundation for
the predictability (or lack-thereof) of future solar activity. These considerations
motivate the current study.

The main flux transport processes in the SCZ involve magnetic buoyancy (timescale on
the order of months), meridional circulation, diffusion and downward flux-pumping
(timescales relatively larger). Because magnetic buoyancy, i.e., the buoyant rise of
magnetic flux tubes, acts on timescales much shorter than the solar cycle timescale,
the fluctuations that it produces are also short-lived in comparison. Our focus here is
on longer-term fluctuations, on the order of the solar cycle period, that may lead
to predictive capabilities.

Through an analysis of observational data, \citet{hathaway2003} have shown that the
solar cycle amplitude and duration are correlated with the equatorward drift velocity
of the sunspot belts during the cycle. They associate this drift velocity with the deep
meridional counterflow that must exist to balance the poleward flows that are observed
at the surface (\citealp{hathaway1996}, \citeyear{hathaway2005}; \citealp{miesch2005}).
The results show a significant negative correlation between the drift velocity and the
cycle duration, so that the drift is faster in shorter cycles, consistent with the
interpretation of meridional circulation as the timekeeper of the solar cycle
(\citealp{nandy2004}; but see also \citealp{schuessler2004}). In addition
\citet{hathaway2003} identified positive correlations between the drift velocity of
cycle $n$ and the amplitudes of both cycles $n$ and $n+2$. While the two-cycle time lag
was a new result, the positive correlation between circulation speed and amplitude of
the same cycle is supported by several earlier studies. In their surface flux transport
model, \citet{wang2002b} needed a varying meridional flow, faster in higher-amplitude
cycles, to sustain regular reversals in the Sun's polar field. They cited observational
evidence from polar faculae counts \citep{sheeley1991}, which peaked early for two of
the stronger cycles, coinciding with poleward surges of magnetic flux. Furthermore,
observations show a statistically-significant negative correlation between peak sunspot
number and the duration of cycles 1 to 22 (Figure 1c of \citealp{charbonneau2000}; see
also \citealp{solanki2002}). Such a negative correlation between cycle amplitude and
duration is also found in the models of \citet{hoyng1993} and \citet{charbonneau2000}.
Taken with the inverse relation between cycle duration and circulation speed, this is
again suggestive of a positive correlation between circulation speed and cycle
amplitude.

Meridional circulation plays an important role in a certain class of theoretical solar
cycle models often referred to as ``flux-transport'', ``advection-dominated,'' or
``circulation-dominated'' dynamo models (see, e.g., the review by Nandy 2004). Such
models have gained popularity in recent years owing to their success in reproducing
various observed features of the solar cycle
\citep{choudhuri1995,durney1995,dikpati1999,kueker2001,bonanno2002,
nandy2001,nandy2002,chatterjee2004}. In these models, a single-cell meridional
circulation in each hemisphere (which is observed at the solar surface) is invoked to
transport poloidal field, first poleward at near-surface layers and then down to the
tachocline where toroidal field is generated. Subsequently, the return flow in the
circulation advects this toroidal field belt equatorward through a region at the base
of the SCZ which is characterized by low diffusivity. From this deep toroidal field
belt, destabilized flux tubes rise to the surface due to magnetic buoyancy, producing
sunspots \citep{parker1955}. We may point out here that the name
``flux-transport dynamo'' is somewhat inappropriate to classify a circulation or
advection-dominated dynamo (where the diffusion timescale is much larger than the
circulation timescale throughout the dynamo domain). Our results indicate that
diffusive flux-transport in the SCZ could play a dominant role in dynamos even when the
cycle period is governed by meridional circulation speed, pointing out that
flux-transport is a shared process. So, henceforth, by ``flux-transport'' dynamo, we
imply a dynamo where the transport of magnetic field is shared by magnetic buoyancy,
meridional circulation, and diffusion.

Flux-transport dynamos offer the possibility of prediction because of their inherent
memory. This arises specifically when the dynamo source regions for poloidal field
production (the traditional $\alpha$-effect) and toroidal field generation (the
$\Omega$-effect) are spatially segregated. A brief discussion on important timescales
(we identify three significant ones) in the dynamo process is merited here. The first
is governed by the buoyant rise of toroidal flux tubes from the $\Omega$-effect layer
to the $\alpha$-effect layer to generate the poloidal field; since this is a fast
process on the order of months, no significant memory is introduced here. The second
involves the transport of poloidal field back into the $\Omega$-effect layer (either by
circulation or diffusion). This could be a slow process where significant memory is
introduced which is dominated by the fastest of the competing processes (advection
versus diffusion). The third timescale relates to the slow equatorward transport of the
toroidal field belt through the base of the SCZ, which sets the period of the sunspot
cycle. In this class of dynamo models, with meridional circulation and low
diffusivities in the tachocline (at the base of the SCZ), the third timescale is almost
invariably determined by the circulation speed. It is the second timescale above, with
competing effects of diffusive flux transport and advective flux transport, that becomes
important in the context of the persistence of memory. In the advection-dominated,
stochastically fluctuating model of \citet{charbonneau2000}, this second timescale
(governed by advection of poloidal field due to meridional circulation) was about 17
years, so that the polar field at the end of cycle $n$ correlated strongest with the
toroidal field of cycle $n+2$ rather than that of cycle $n+1$. The length of memory of
any particular flux-transport dynamo model is unfortunately dependent on the internal
meridional flow profile, and on other chosen properties of the convection zone which
are not yet well-determined observationally. A particular problem is the strength of
diffusivity in the convection zone, which strongly affects the mode of operation of the
dynamo.

Even if one assumes that these flux-transport dynamos capture enough of the realistic
physics of the SCZ to make predictions of future solar activity, these predictions are
critically dependent on the relative role of diffusion and advection in the SCZ.
\citet{dikpati2006b}, in their highly {\it advection-dominated} model, show that bands
of latitudinal field from three previous cycles remain ``lined up in the meridional
circulation conveyor belt''. They suggest that poloidal fields from cycles $n-3$,
$n-2$, and $n-1$ combine to produce the toroidal field of cycle $n$. Based on an
assumed proxy for the solar poloidal fields (sunspot area), this leads them to predict
that Cycle 24 will be about $50\%$ stronger than Cycle 23 \citep{dikpati2006}. In stark
contrast, \citet{choudhuri2007}, using a flux-transport dynamo model with {\it
diffusion-dominated} SCZ, and using as inputs the observed strength of the solar dipole
moment (as a proxy for the poloidal field), predict that Cycle 24 will be about $35\%$
\emph{weaker} than Cycle 23. \citet{choudhuri2007} argue that the main contribution to
the toroidal field of cycle $n$, comes only from  the polar field of cycle $n-1$ (see
also \citealp{jiang2007} for further details of this model).

The conflicting predictions from these two solar dynamo models presumably result from
the difference in the memory (i.e., survival) of past cycle fields in these models and
could be to some extent influenced by the different inputs they use as proxies for the
solar poloidal field. We also hypothesize that stronger diffusion in the
\citet{choudhuri2007} model destroys polar field faster, and that flux transport by
diffusion across the SCZ in this model short-circuits the meridional circulation
conveyor belt, thereby shortening the memory of previous cycles. We perform a detailed
analysis to test these ideas. To begin with, we consider a wider parameter space in the
present paper, where we study the effect of varying meridional circulation speed and
SCZ diffusivity on the amplitude and period of the solar cycle. In these simulations,
we keep all other parameters the same, allowing a direct comparison between advection-dominated and diffusion-dominated SCZ regimes -- which has previously been clouded by
other differences between models. Then we introduce stochastic fluctuations in the
model $\alpha$-effect to self-consistently generate cycle-amplitude variations -- as a
completely theoretical construct towards studying cycle-to-cycle variations, in
contrast to using diverse observed proxies for time-varying poloidal fields.
Subsequently, we perform a comparative analysis of the persistence of memory in this
stochastically forced dynamo model in both the advective and diffusive flux-transport
dominated regimes. Therefore, in spirit, this paper deals with the underlying physics
of solar cycle predictability, and is not concerned with making a prediction itself.
The layout of this paper is as follows. The main features of the model are summarised
in Section \ref{sec:model}, and the results of the parameter-space study are presented
in Section \ref{sec:results}. These results are interpreted in Section
\ref{sec:regimes}. In Section \ref{sec:timedelays} we analyze the persistence of memory
in the advection versus diffusion dominated regimes. We conclude in Section
\ref{sec:conclusion} with a discussion on the relevance of this work in the context of
developing predictive capabilities for the solar activity cycle.

\section{The Model} \label{sec:model}

We use the solar dynamo code \emph{Surya}, which has been studied extensively in
different contexts (e.g. \citealp{nandy2002}, \citealp{chatterjee2004},
\citealp{chatterjee2006}), and is made available to the public on request. The major
ingredients of the code include an analytic fit to the helioseismically-determined
differential rotation profile, a single-cell meridional circulation in the SCZ,
different diffusivities for the toroidal and poloidal fields, a buoyancy algorithm to
model radial transport of magnetic flux, and a Babcock-Leighton (BL;
\citealp{babcock1961}, \citealp{leighton1969}) type $\alpha$-effect localized near the surface layer (signifying the
generation of poloidal field due to the evolution of
tilted bipolar sunspot pairs under surface flux transport). The code solves the
kinematic mean-field dynamo equations for an axisymmetric magnetic field, which may be
expressed in spherical coordinates $(r,\theta,\phi)$ as
\begin{equation}
\mathbf{B} = B(r,\theta)\mathbf{e}_{\phi} + \bpol,
\end{equation}
where $B(r,\theta)$ and $\bpol=\nabla\times\left [ A(r,\theta)\mathbf{e}_{\phi} \right ]$
correspond to the toroidal and poloidal components respectively. The mean-field MHD
induction equation (see e.g. \citealp{moffatt1978}) then leads to the following
standard equations for the $\alpha$-$\Omega$ dynamo problem:
\begin{align}
\pdiff{A}{t} + \frac{1}{s}\left( \mathbf{v}\cdot\nabla \right)\left(sA\right) &= \eta_p\left(\nabla^2 - \frac{1}{s^2}\right)A + \alpha B,  \label{eqn:Aevol}\\
\pdiff{B}{t} + \frac{1}{r}\left( \pdiff{}{r}\left( r v_r B \right) + \pdiff{}{\theta}\left( v_\theta B\right) \right) &= \eta_t \left(\nabla^2 - \frac{1}{s^2} \right)B \nonumber \\
 &+ s\left( \bpol \cdot \nabla \right) \Omega + \frac{1}{r}\frac{d\eta_t}{dr}\pdiff{}{r}\left( rB \right). \label{eqn:Bevol}
\end{align}
Here $s=r\sin\theta$,  and we specify the meridional flow $\mathbf{v}$, the internal
angular velocity $\Omega$, the diffusivities $\eta_p$ and $\eta_t$, and the coefficient
$\alpha$ for the BL $\alpha$-effect which describes the generation of poloidal field at the solar
surface from the decay of bipolar sunspots. Note that although for modelling purposes
the BL $\alpha$-effect is mathematically similar to the traditional mean-field
$\alpha$-effect due to small-scale helical turbulence, the former is fundamentally
different. The BL $\alpha$-effect acts on much larger spatial (on the order of active
regions or greater) and temporal (surface flux transport) scales, and is quenched at
much stronger field strengths ($10^5 \mathrm{G}$). The profiles of $\Omega$ and $\alpha$ were
described in \citet{chatterjee2004} and will not be repeated here. We will, however,
describe the meridional circulation and diffusivity profiles in more detail.

\subsection{Meridional Circulation}

The meridional circulation is defined in terms of a streamfunction $\psi(r,\theta)$,
giving the velocity by
\begin{equation}
\rho\mathbf{v}=\nabla\times\left [\psi(r,\theta)\mathbf{e}_\phi\right ],
\end{equation}
where we assume the density stratification
\begin{equation}
\rho=C\left(R_\odot/r - 0.95\right)^{3/2}.
\end{equation}
The streamfunction is given by
\begin{align}
\psi r\sin\theta &= \psi_0\left(r-R_p\right)\sin\left(\frac{\pi(r-R_p)}{R_\odot-R_p}\right) \nonumber\\
& \qquad \times
\left\{1-\textrm{e}^{-\beta_1\theta^\epsilon}\right\}\left\{1-\textrm{e}^{\beta_2(\theta-\pi/2)}\right\}\textrm{e}^{-\left((r-r_0)/\Gamma\right)^2},
\end{align}
where $\beta_1=1.5\times 10^{-8}\permet$, $\beta_2=1.8\times
10^{-8}\permet$, $\epsilon=2.0000001$, $\Gamma=3.47\times 10^{8}\met$, and
$r_0=(R_\odot-R_b)/4$. Here $R_\odot=6.96\times 10^{8}\met$ is the radius of the Sun,
$R_b=0.55R_\odot$ is the bottom of the simulation domain, and $R_p=0.61R_\odot$ is the
penetration depth of the meridional circulation. We combine the arbitrary constants $C$
and $\psi_0$ in the parameter $v_0=-\psi_0/(0.95C)$ which gives, approximately, the
flow speed near the surface at mid-latitudes. It is this parameter $v_0$ which we vary
to change the circulation speed in this study.

\begin{figure}
\plottwo{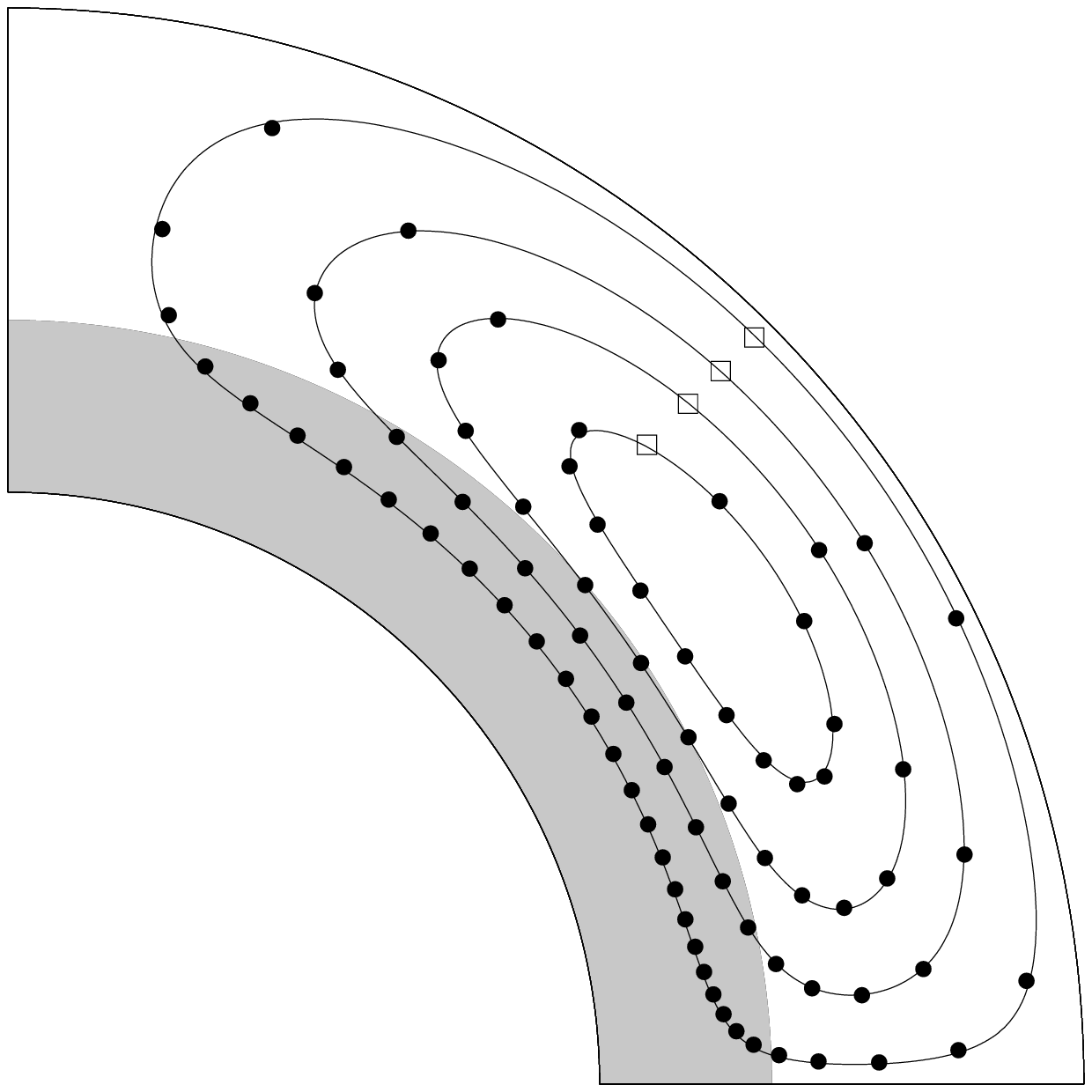}{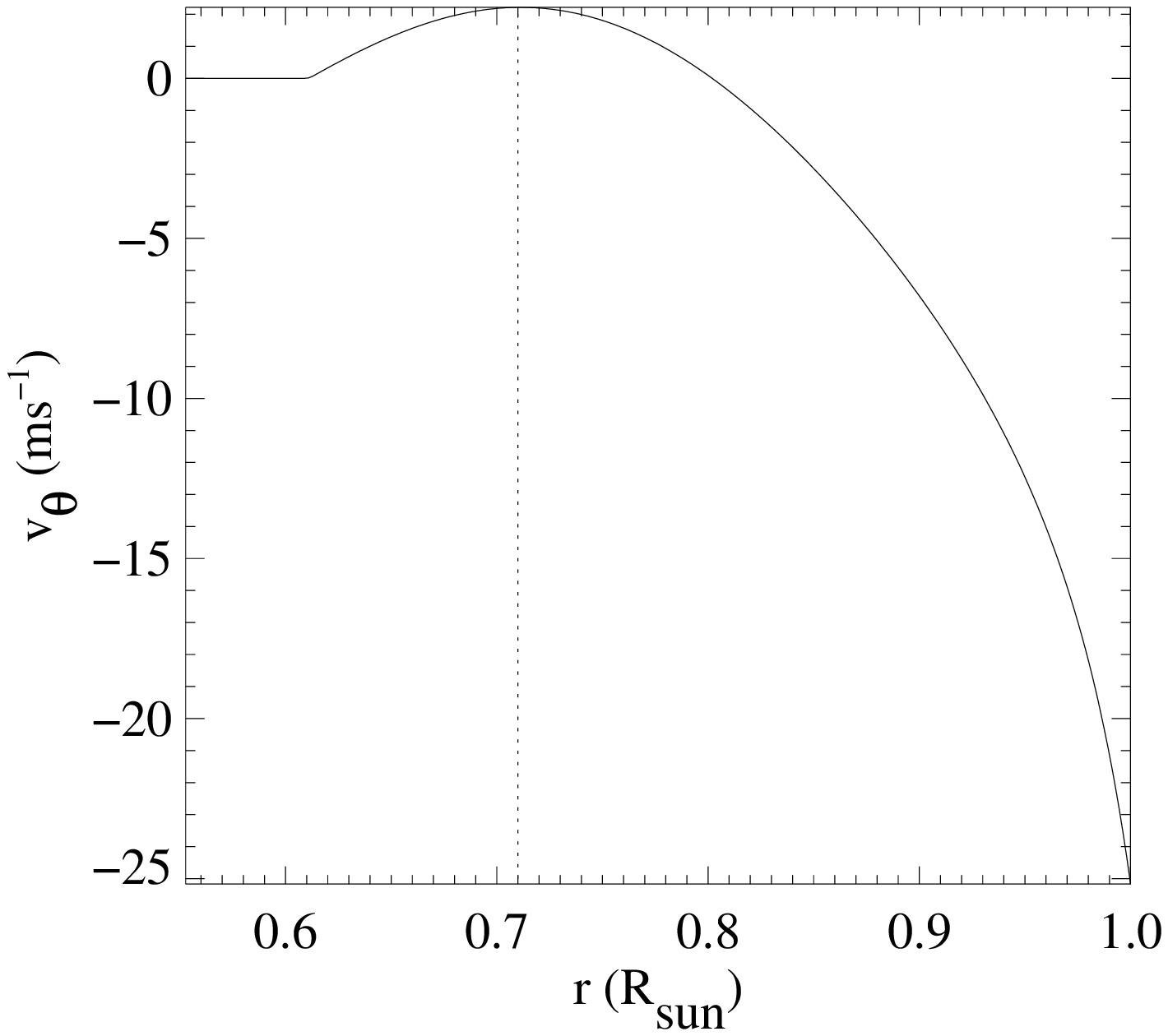}
 \caption{Streamlines of the meridional circulation profile in our model (\emph{left}), and latitudinal velocity profile across a radial cut at $\theta=45^\circ$ (\emph{right}). Dots on the streamlines show yearly positions for particles with $v_0=25\mpsec$, starting from the squares at $\theta=45^\circ$ and moving anti-clockwise.} \label{fig:merid}
\end{figure}

The circulation profile is illustrated in Figure \ref{fig:merid} for $v_0=25\mpsec$.
The dots are plotted at yearly intervals for particles moving along the streamlines
shown.

\subsection{Diffusion}

\begin{figure}
\epsscale{0.7}
\plotone{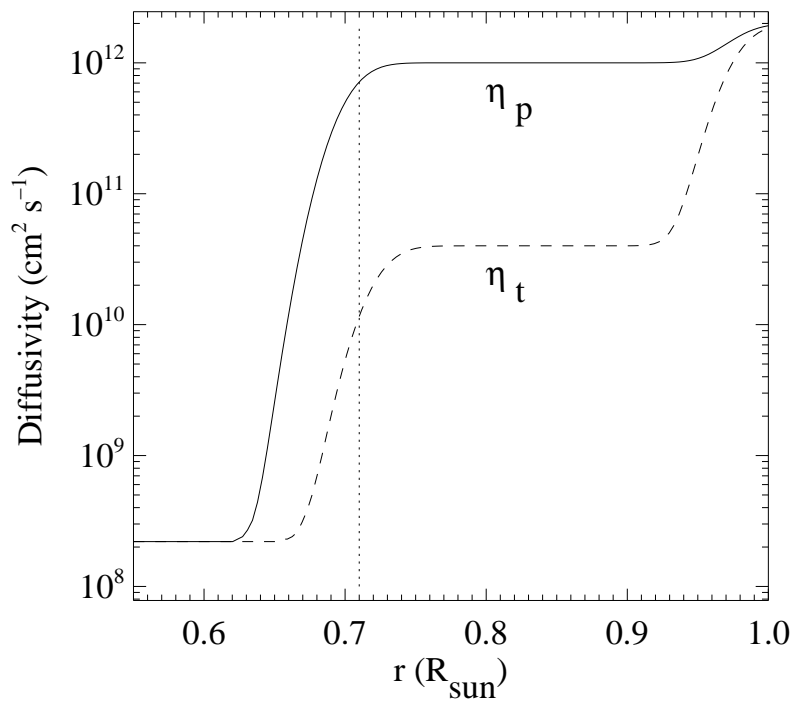}
\caption{Example
diffusion profiles as a function of $r$ for the toroidal ($\eta_t$) and poloidal
($\eta_p$) fields. The dotted line shows the location of the tachocline. Here
$\eta_0=2.0\times 10^{12}\cmsqpsec$, $\eta_1=0.04\times 10^{12}\cmsqpsec$, and
$\eta_2=1.0\times 10^{12}\cmsqpsec$.} \label{fig:eta}
\end{figure}

We use different diffusivities for the toroidal and poloidal fields, defined as
follows:
\begin{align}
\eta_t(r) &= \eta_{RZ} + \frac{\eta_1-\eta_{RZ}}{2}\left[1+\erf\left(\frac{r-r'_{BCZ}}{d_t}\right)\right]\nonumber\\
& \qquad\qquad\qquad +\frac{\eta_0-\eta_1}{2}\left[1+\erf\left(\frac{r-r_{TCZ}}{d_t}\right)\right],\\
\eta_p(r) &= \eta_{RZ} + \frac{\eta_2-\eta_{RZ}}{2}\left[1+\erf\left(\frac{r-r_{BCZ}}{d_t}\right)\right] \nonumber\\
& \qquad\qquad\qquad +
\frac{\eta_0-\eta_2}{2}\left[1+\erf\left(\frac{r-r_{TCZ}}{d_t}\right)\right].
\end{align}
Here $d_t=0.025R_\odot$, $r_{BCZ}=0.7R_\odot$, $r'_{BCZ}=0.72R_\odot$, and
$r_{TCZ}=0.975R_\odot$. In the radiative core we choose a low diffusivity, namely
$\eta_{RZ}=2.2\times 10^8\cmsqpsec$, representing molecular diffusivity only since
there is no turbulent convection. We always choose $\eta_1<\eta_2$ so that the toroidal
field diffusivity $\eta_t$ in the convection zone is lower than the poloidal field
diffusivity $\eta_p$. This is to model the suppression of turbulent diffusivity by
strong magnetic fields, as toroidal field tends to be strong and concentrated in
localised flux tubes and therefore subject to less diffusion \citep{choudhuri2003},
whereas poloidal field is weaker and subject to more diffusion. At the surface both
diffusivities increase to a high value (of the order of $10^{12}\cmsqpsec$), in line
with surface flux transport models and observational estimates. Typical profiles are
illustrated in Figure \ref{fig:eta}.

\subsection{Numerical Domain and Boundary Conditions}

We solve Equations \eqref{Aevol} and \eqref{Bevol} in a meridional plane $0.55 R_\odot
< r < R_\odot$, $0 < \theta < \pi/2$, representing the Northern hemisphere. This is
divided into a spherical grid of 128 by 128 cells, uniformly spaced in $r$ and
$\theta$. We use the same boundary conditions as \citet{chatterjee2004}, except that we
consider only the Northern hemisphere and set $B=0$ and $\pdlin{A}{\theta}=0$ at the
equator ($\theta=\pi/2$), thereby forcing the solution to have dipolar parity.

\subsection{Example Solutions}

\begin{figure}
\epsscale{0.7}
\plotone{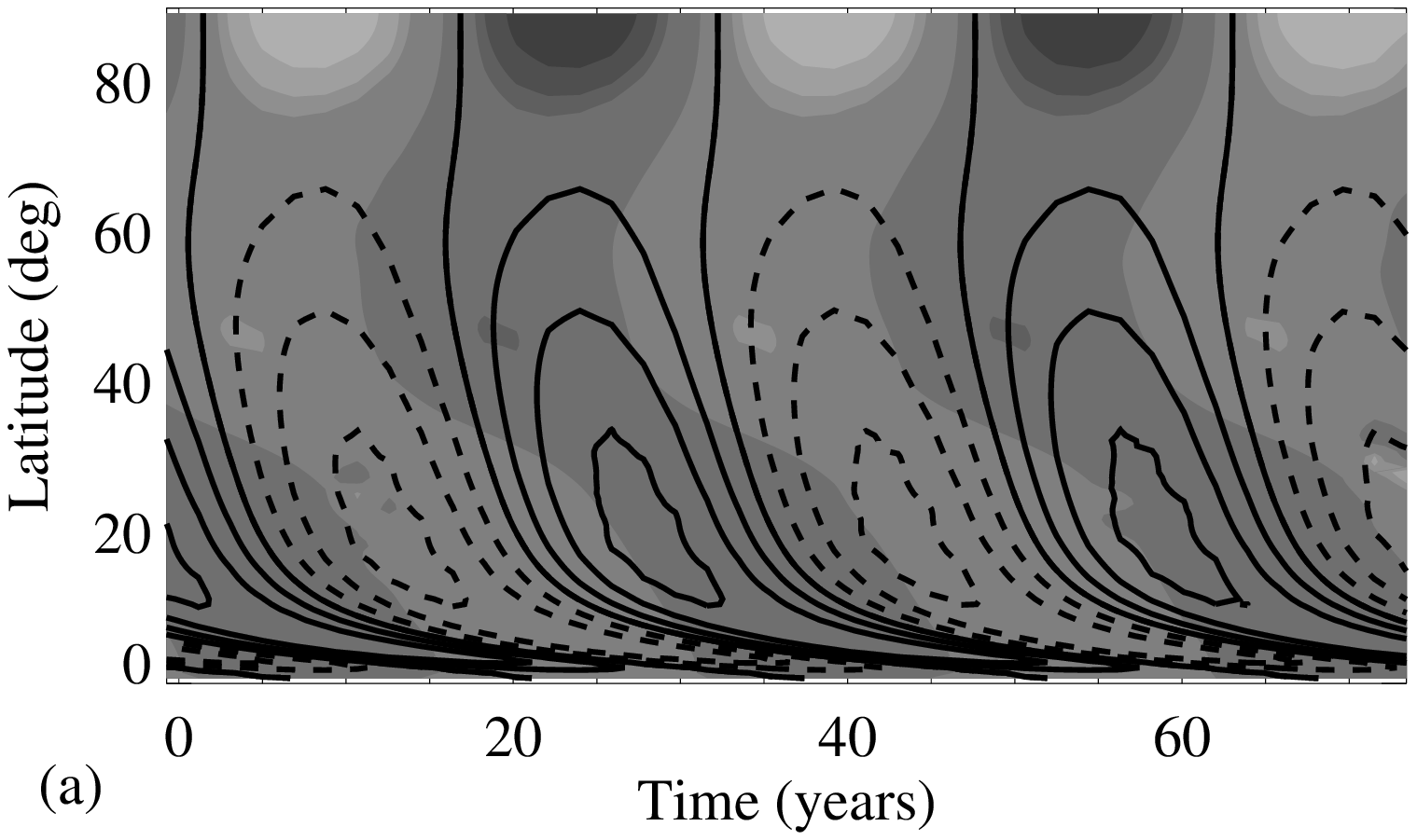}
\plotone{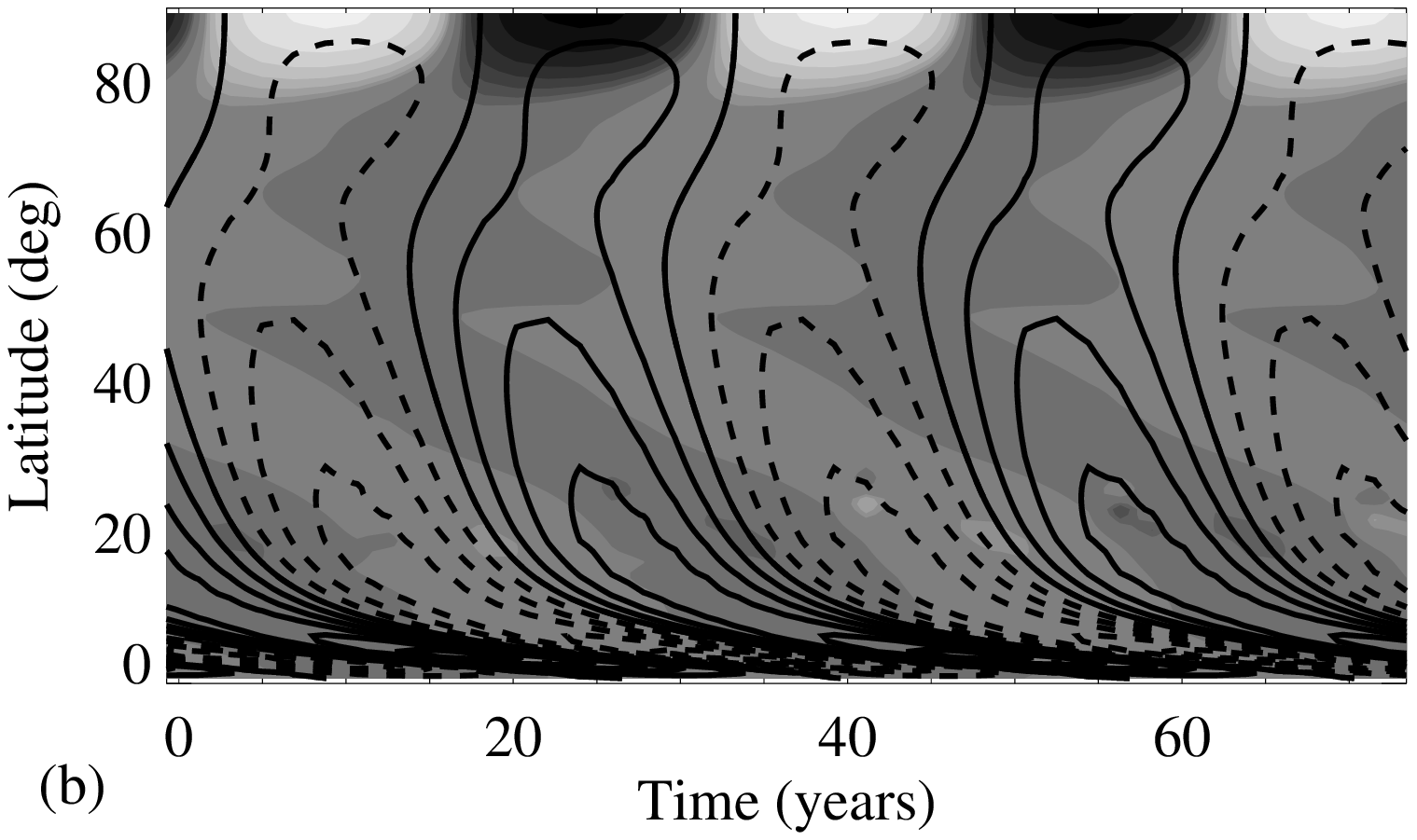}
\caption{Two example
solutions: (a) with $v_0=20\mpsec$ and $\eta_2=2.0\times10^{12}\cmsqpsec$
(characterizing a diffusive flux-transport dominated SCZ); and (b) with the same $v_0$
but $\eta_2=0.5\times10^{12}\cmsqpsec$ (characterizing an advective flux-transport
dominated SCZ). In each case black lines are contours of toroidal field $B$ at the base
of the convection zone (solid lines for positive values, dashed for negative). The
grayscale in the background shows surface radial field strength $B_r(r=R_\odot)$, with
white for positive and black for negative. The same contour levels are used in both
plots.} \label{fig:fields}
\end{figure}

Example solutions for two different runs are shown in time-latitude plots in Figures
\ref{fig:fields}(a) and (b), where the black lines denote contours of toroidal field
strength at the base of the convection zone ($r=0.71R_\odot$). This corresponds to the
solar butterfly diagram, with the strongest field located at the active latitudes and
migrating equatorward during each cycle. The background shading shows the strength of
the radial field at the solar surface ($r=R_\odot$), which peaks at the pole several years after the toroidal field maxima (of the same sign) at low latitudes. The two solutions in Figure \ref{fig:fields} characterize the diffusion-dominated SCZ (Figure \ref{fig:fields}a) and advection-dominated SCZ (Figure \ref{fig:fields}b) regimes of the dynamo. In Figure \ref{fig:fields}(b) the toroidal field shows a poleward branch at high latitude which is absent in Figure \ref{fig:fields}(a), and also a stronger radial polar field at the surface. The cause of these differences between the two regimes will become clear in Section \ref{sec:regimes}.


\section{Results} \label{sec:results}

We have carried out a parameter-space study to investigate how the cycle duration and
amplitude in our model depend on the speed of meridional circulation and on the
diffusivity in the convection zone. In each run of the code the parameters are held
constant in time, but they are varied between different runs. Specifically, we vary the
parameter $v_0$, which gives the maximum circulation speed, and also $\eta_2$, which
affects the diffusive decay and transport of the \emph{poloidal} field in the
convection zone, but not the toroidal field. In all runs we keep a surface diffusivity
of $\eta_0 = 2.0 \times 10^{12}\cmsqpsec$, and a toroidal field diffusivity of $\eta_1
= 0.04 \times 10^{12}\cmsqpsec$ in the convection zone. These choices approximate the
fact that turbulent diffusion is expected to be more efficient in the decay and
dispersal of the weaker poloidal field, but less so for the stronger toroidal field;
the latter suppresses the convective motions that give rise to turbulent diffusivity in
the first place. The $\alpha$-effect coefficient is not varied, but is set to $\alpha_0
= 30 \mpsec$ for each run. This particular value was chosen to ensure that periodic
solutions could be obtained for a wide range of the parameters $v_0$ and $\eta_2$. Each
run was started from an arbitrary initial state, and then evolved until initial
transients had disappeared, leaving a steady periodic dynamo solution. Such a periodic
solution was found to exist only within a certain range of $v_0$ for each value of
$\eta_2$. The cycle duration and amplitude were then measured from the periodic
solutions, in the cases where such a solution was found.

We define the cycle duration and amplitude by considering the time evolution of the
toroidal magnetic flux $\Phi_\mathrm{tor}$ in a certain region around the base of the
convection zone. Specifically, the toroidal field $B(r,\theta)$ is integrated at each
time step over a region $r=0.677 R_\odot$ to $0.726 R_\odot$, $\theta=45^\circ$ to
$80^\circ$ (i.e., over the tachocline and latitudes $45^\circ$ to $10^\circ$). This
magnetic flux $\Phi_\mathrm{tor}$ should be proportional to the active region magnetic
flux at the solar surface, under the assumption that more toroidal flux at the base of
the convection zone leads to more buoyant eruptions. In a steady dynamo solution the
flux $\Phi_\mathrm{tor}$ varies in strictly periodic manner, with its maximum amplitude
giving the ``cycle amplitude''. We define the ``cycle duration'' to be the interval
between successive peaks of $|\Phi_\mathrm{tor}|$ (half of the full dynamo period).
This is therefore equivalent to the standard definition of the 11-year solar activity
cycle, but of course the simulated periods are different.

The resulting cycle duration and cycle amplitude are plotted as functions of the
circulation speed $v_0$ in Figures \ref{fig:dur_v0} and \ref{fig:amp_v0} respectively.
In these figures each curve corresponds to a different value of the diffusivity
$\eta_2$. The range of speeds covered by each curve indicates the range for which the
code relaxed to a steady periodic dynamo solution, up to a maximum of $v_0=38\mpsec$.
In Figure \ref{fig:amp_diff} the cycle amplitude is plotted as a function of $\eta_2$,
and in this case each curve corresponds to a different circulation speed $v_0$.


\subsection{Dependence of Cycle Period on Meridional Circulation and Diffusion}

\begin{figure}
\plotone{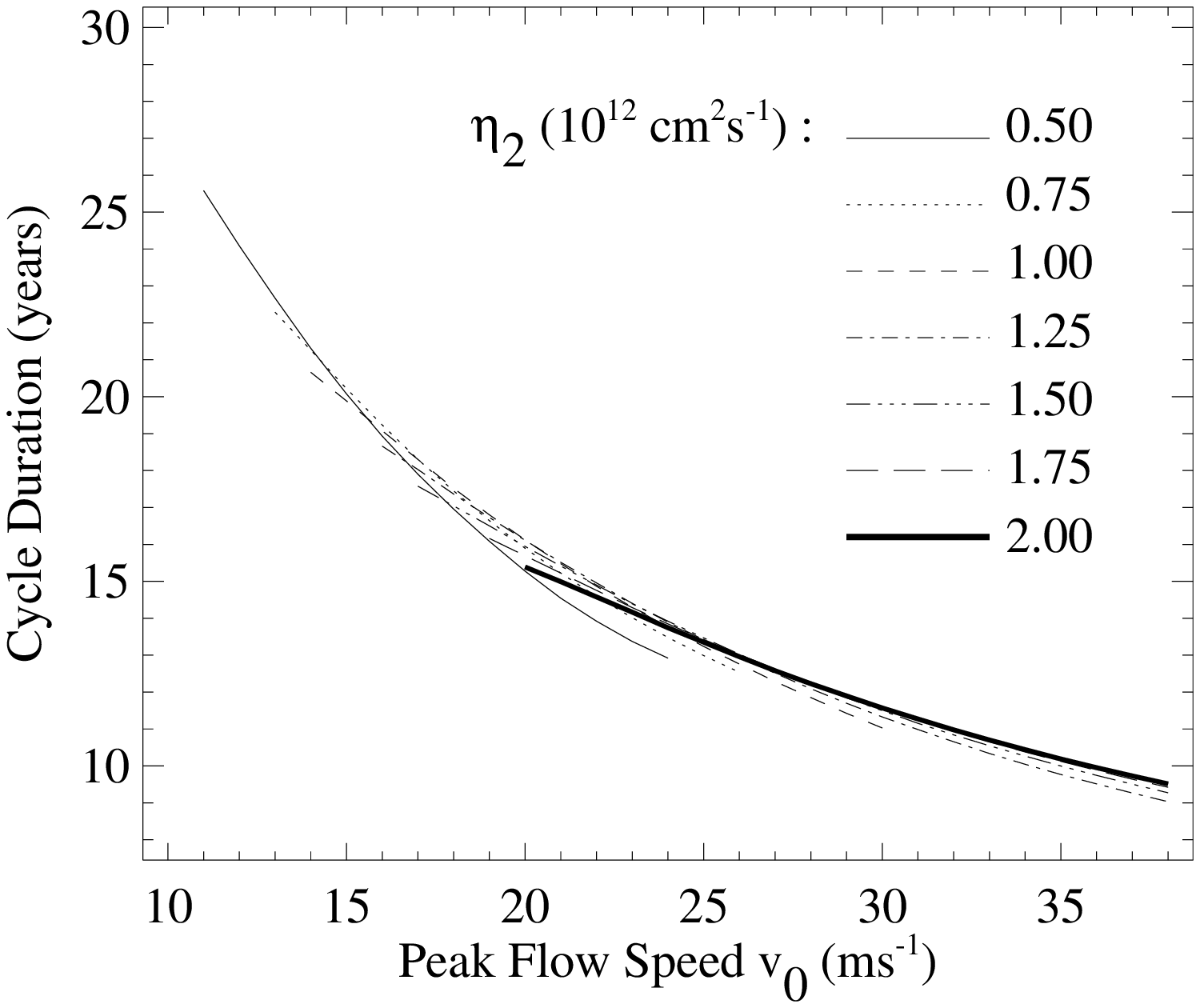}
\caption{Dependence of
cycle duration on the meridional circulation speed $v_0$. Each line style corresponds
to a different value of $\eta_2$ (the poloidal diffusivity in the convection zone) as
given in the legend.} \label{fig:dur_v0}
\end{figure}

Figure \ref{fig:dur_v0} shows a clear inverse dependence of the cycle duration on the
meridional circulation speed $v_0$, with faster circulation leading to shorter cycles.
A least-squares fit for the curve with $\eta_2=0.5\times 10^{12}\cmsqpsec$ gives the
dependence of the cycle period on meridional flow speed
\begin{equation}
T = 217.716\,v_0^{-0.885},
\end{equation}
where $T$ is in years and $v_0$ is in metres per second. This agrees with the $T \sim
v_0^{-0.89}$ found by \citet{dikpati1999}, and this inverse relation is a
well-established result for Babcock-Leighton dynamo models. In these models the
circulation, and specifically the equatorward counterflow at the bottom of the
convection zone, is the primary determinant of the cycle period \citep{nandy2004}.

The cycle duration is only weakly dependent on the diffusivity $\eta_2$. A power-law
fit for the curve with $\eta_2=2.0\times 10^{12}\cmsqpsec$ gives $T =
150.745\,v_0^{-0.756}$ years. The lower power of $v_0$ here indicates that for
higher-diffusivity solutions the cycle duration is slightly less dependent on
circulation speed, presumably because flux transport by diffusive dispersal starts
becoming important. Also, it is evident from Figure \ref{fig:dur_v0} that, at lower
circulation speeds, there is a maximum diffusivity for which a periodic solution can
exist. If there is too much diffusion at a low circulation speed, then the poloidal
field will decay too much during its transport from high to low latitudes, thus
generating insufficient toroidal field to sustain a periodic dynamo process. The
essential difference between advective and diffusive flux transport is that the latter
also reduces field strength during transport, due to diffusive decay.

\subsection{Dependence of Cycle Amplitude on Meridional Circulation and Diffusion}

\begin{figure}
\plotone{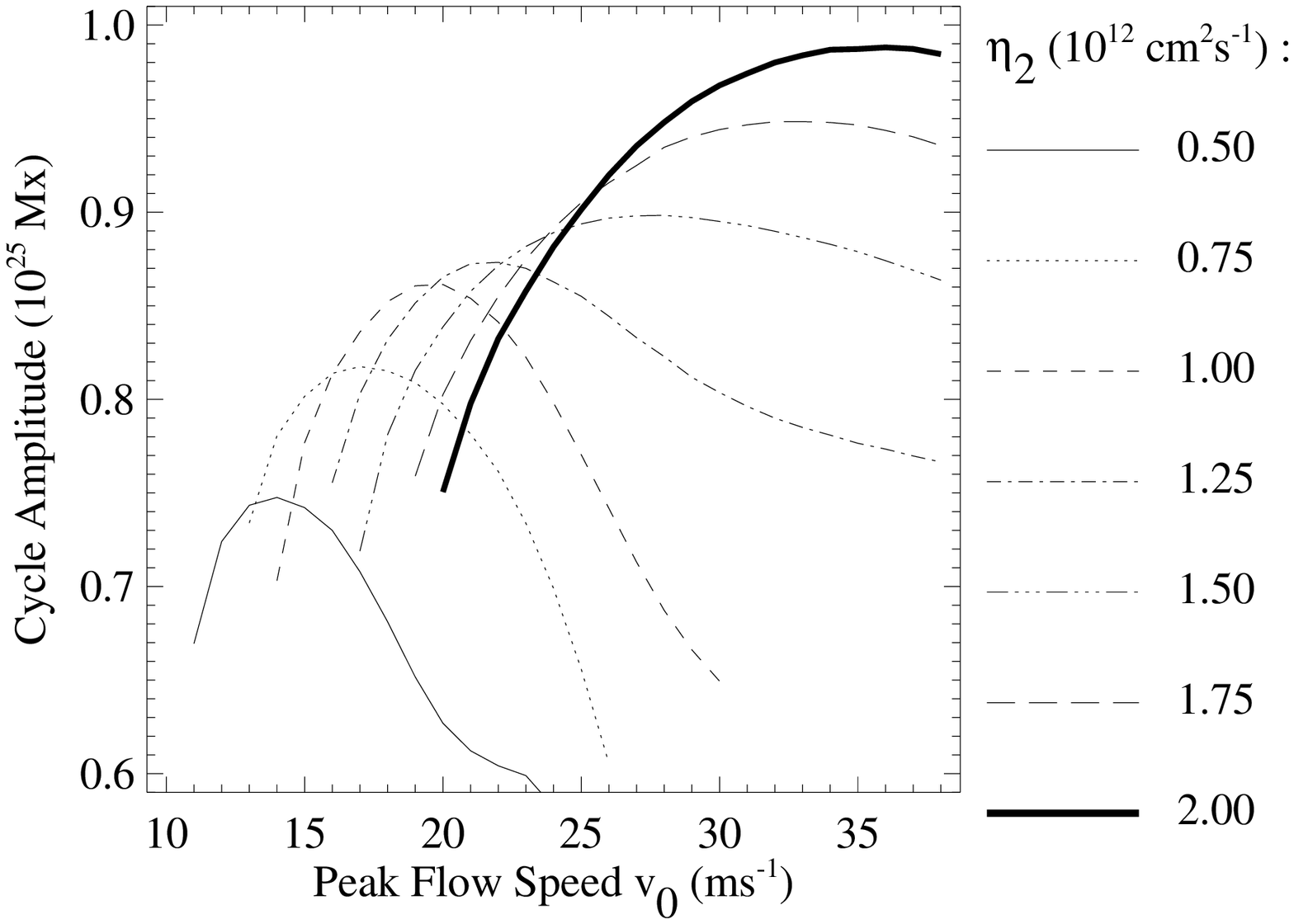}
\caption{Dependence of
cycle amplitude on the meridional circulation speed $v_0$. Each line style corresponds
to a different value of $\eta_2$ (the poloidal diffusivity in the convection zone) as
given in the legend.} \label{fig:amp_v0}
\end{figure}

Now we turn to the dependence of cycle amplitude on the speed of meridional
circulation. This is shown in Figure \ref{fig:amp_v0}, where each curve corresponds to
a different diffusivity $\eta_2$ according to the legend. Rather than being monotonic,
the cycle amplitude first increases with $v_0$ for low $v_0$ and then decreases with
$v_0$ for large $v_0$, with a turnover at some value of $v_0$ in between. The location
of this turnover shifts to higher speeds as the diffusivity is increased.

\begin{figure}
\plotone{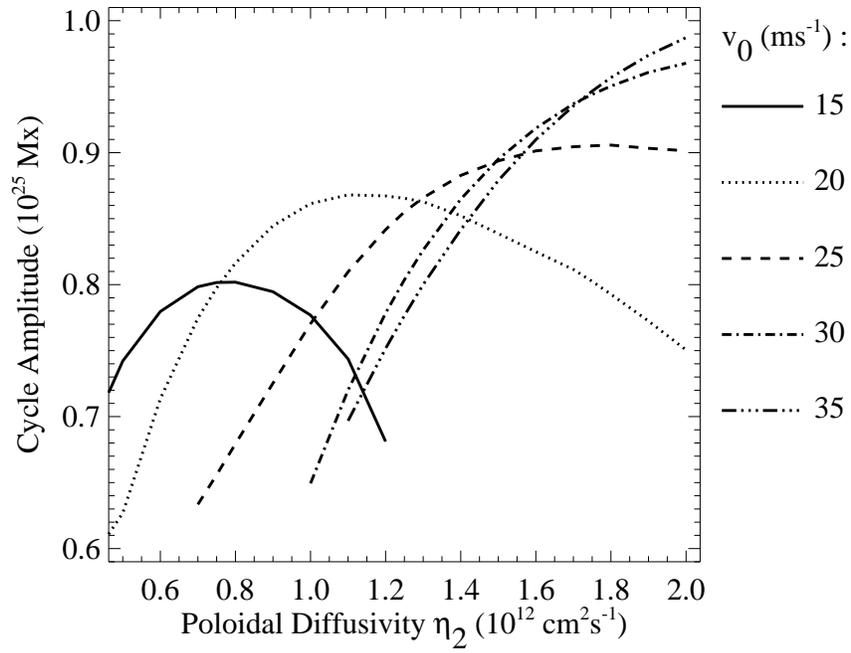}
\caption{Dependence of cycle amplitude on the poloidal diffusivity $\eta_2$ in the
convection zone. Each line style corresponds to a different value of the meridional
circulation speed $v_0$, as given in the legend.} \label{fig:amp_diff}
\end{figure}

The dependence of cycle amplitude on diffusivity at any given circulation speed is not
entirely clear on Figure \ref{fig:amp_v0}, but is evident in Figure \ref{fig:amp_diff},
where cycle amplitude is plotted against diffusivity $\eta_2$. In this figure each
curve corresponds to a different value of the circulation speed $v_0$. We see a similar behavior in that the cycle amplitude first increases with $\eta_2$ for
low $\eta_2$ and then decreases with $\eta_2$ for high $\eta_2$, with a turnover in
between. If the circulation speed is increased, then the value of $\eta_2$
corresponding to this turnover also increases.

The behaviour of the cycle amplitude in our model, as illustrated in Figures
\ref{fig:amp_v0} and \ref{fig:amp_diff}, is more complex than expected \emph{a priori}.
Rather than a simple linear dependence on the circulation speed $v_0$, there is a
turnover in cycle amplitude, at a speed which changes depending on the diffusivity in
the convection zone. In the next section we investigate the cause of this behaviour in
the model.

\section{Advection versus Diffusion Dominated Solar Convection Zones} \label{sec:regimes}

\begin{figure}
\hbox{
\includegraphics[width=0.46\textwidth]{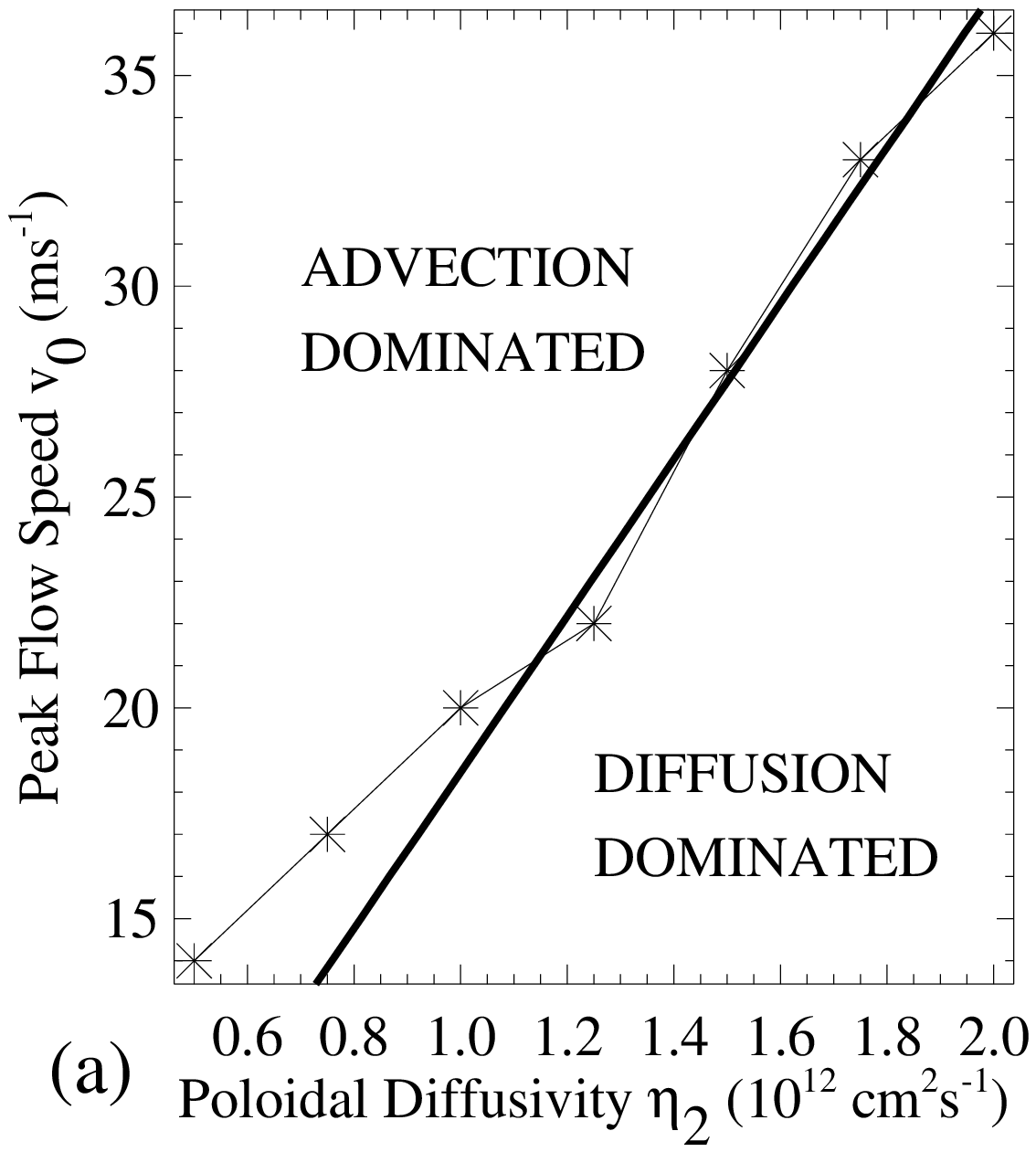}
\includegraphics[width=0.54\textwidth]{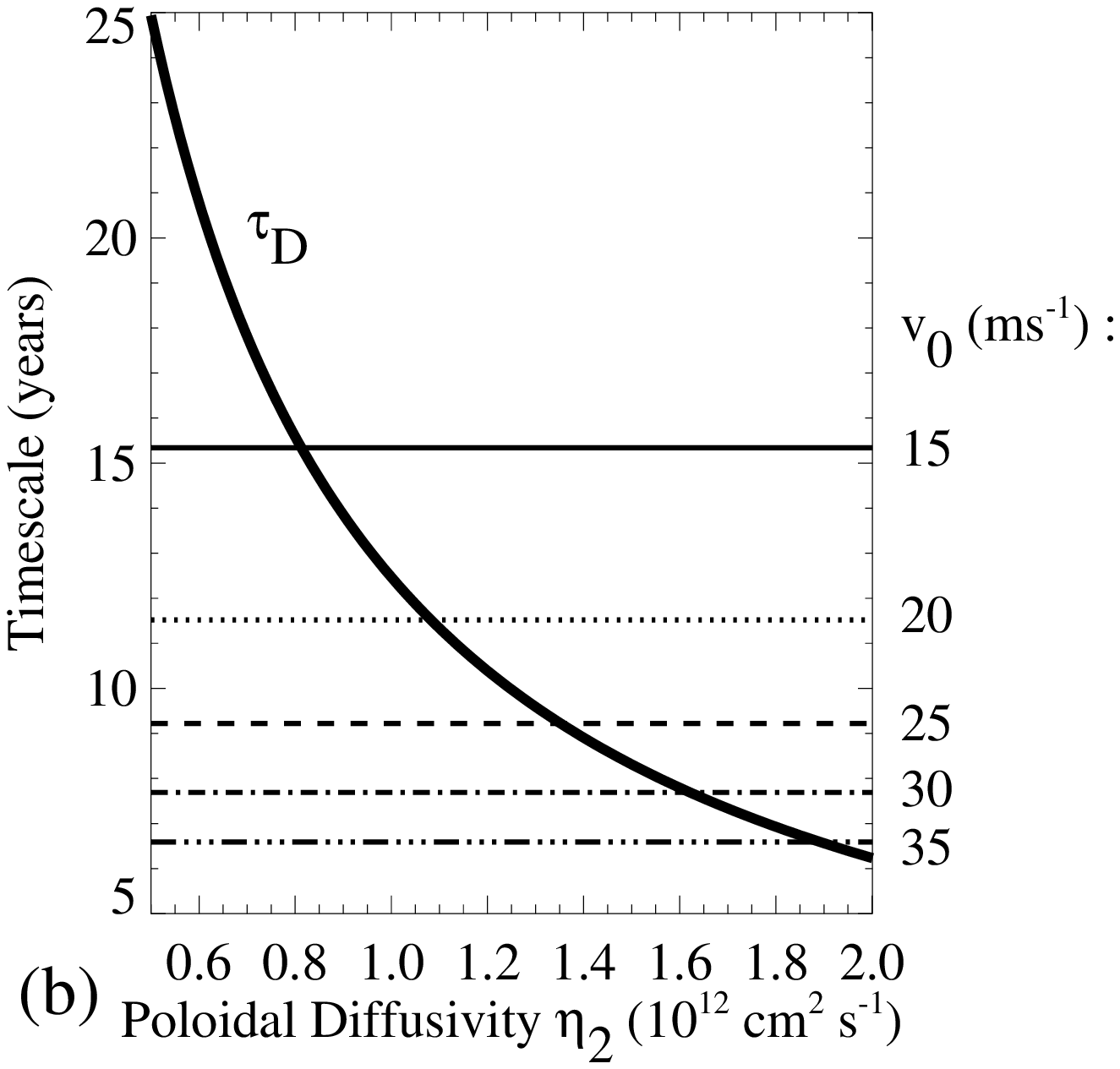}
}
\caption{Transition
between the advection-dominated and diffusion-dominated regimes. In (a) asterisks
indicate the flow speeds $v_0$ corresponding to turnover of cycle amplitude for fixed
values of $\eta_2$ (inferred from the simulations shown in Figures~5 and 6). The bold
line shows the transition point that may be inferred from simple theoretical comparison
of circulation and diffusion timescales. Panel (b) shows the diffusion timescale
$\tau_{\textrm{D}}$ as a function of $\eta_2$ (bold line), and circulation timescales
$\tau_{\textrm{C}}$ for selected speeds $v_0$ (horizontal lines), as defined in the
text.} \label{fig:regimes}
\end{figure}

The turnover of cycle amplitude as depicted in Figure \ref{fig:amp_v0} occurs at a
higher circulation speed $v_0$ as the diffusivity $\eta_2$ is increased. The asterisks
(joined by a thin line) in Figure \ref{fig:regimes}(a) show the location of this
turnover as a function of $\eta_2$. We may think of this line in the $(\eta_2,v_0)$
plane as the dividing line between two distinct regimes of the dynamo, which we call
\emph{advection-dominated} and \emph{diffusion-dominated}. The advection-dominated
regime corresponds to high circulation speed and low diffusivity, while the
diffusion-dominated regime corresponds to low circulation speed and high diffusivity. A
shift from one of these regimes to another affects flux-transport dynamics in a way
that results in contrasting dependence of cycle amplitude on the governing parameters.
Consider how the cycle amplitude varies with $v_0$ for a fixed value of $\eta_2$,
corresponding to a curve on Figure \ref{fig:amp_v0}. In the diffusion-dominated regime,
a higher circulation speed means less time for diffusive decay of the poloidal field
during its transport through the convection zone, leading to more generation of
toroidal field and hence a higher cycle amplitude. In the advection-dominated regime, a
higher circulation speed leads to a lower cycle amplitude because there is less time to
amplify toroidal field in the tachocline (through which magnetic fields are swept
through at a faster speed). It is the balance between these conflicting influences that
leads to a turnover in cycle amplitude at some intermediate circulation speed.


The bold line in Figure \ref{fig:regimes}(a) shows the transition point between the two
regimes that may be inferred from a simple balance between circulation and diffusion
timescales $\tau_{\textrm{C}}$ and $\tau_{\textrm{D}}$. For a given circulation speed
$v_0$, we define the circulation timescale $\tau_{\textrm{C}}$ as the time taken for
meridional circulation to advect poloidal fields from $r=0.95R_\odot, \theta=45^\circ$
to the location where the strongest toroidal field is formed at the tachocline
($\theta=60^\circ$). The diffusion timescale is defined as
$\tau_{\textrm{D}}=L^2/\eta_2$, where  $L=0.285R_\odot$ is the radial distance across the
convection zone from the same starting point. The two timescales are compared in Figure
\ref{fig:regimes}(b), where each horizontal line gives the circulation time
$\tau_{\textrm{C}}$ for a different speed $v_0$, and the bold curve gives
$\tau_{\textrm{D}}$ as a function of $\eta_2$. The crossing points of horizontal lines
with this curve give the transition points between the advection dominated
($\tau_{\textrm{C}} < \tau_{\textrm{D}}$) and diffusion dominated ($\tau_{\textrm{C}} >
\tau_{\textrm{D}}$) regimes from these simple theoretical considerations -- which are in
good agreement with the simulated transition points.

\subsection{Magnetic Field Evolution in Advection versus Diffusion Dominated Regimes}

\begin{figure}
\begin{center}
\epsscale{0.47}
\plottwo{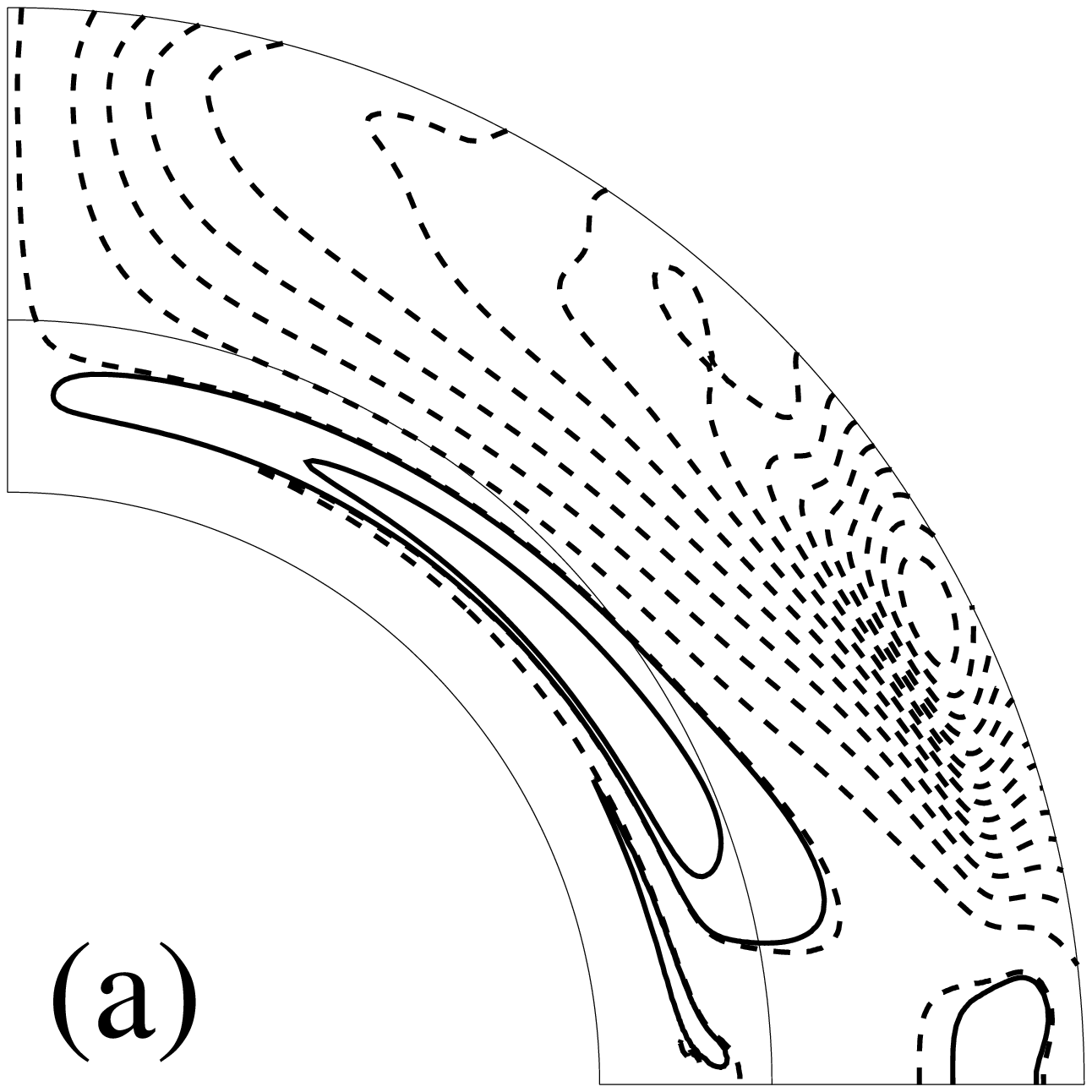}{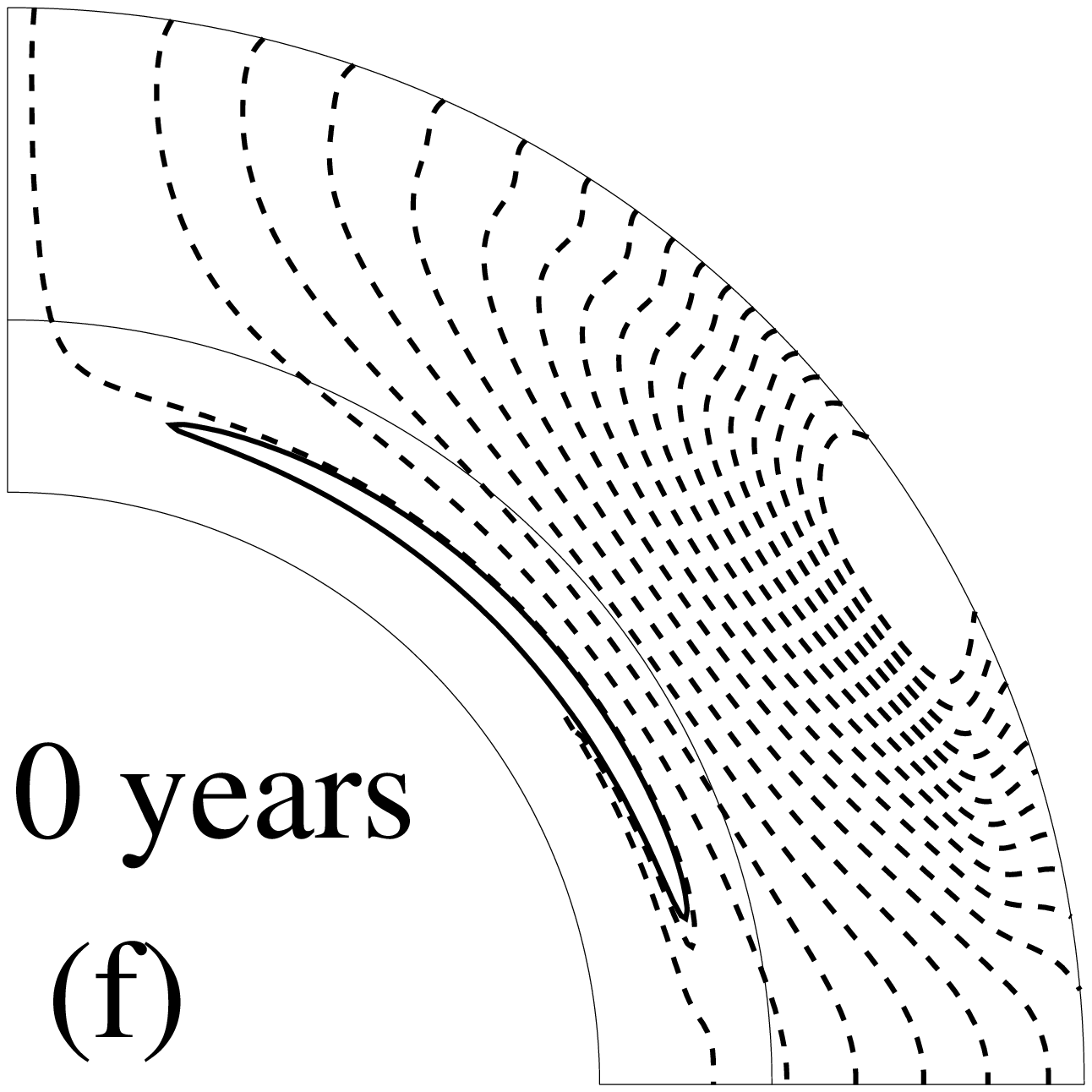}\\
\plottwo{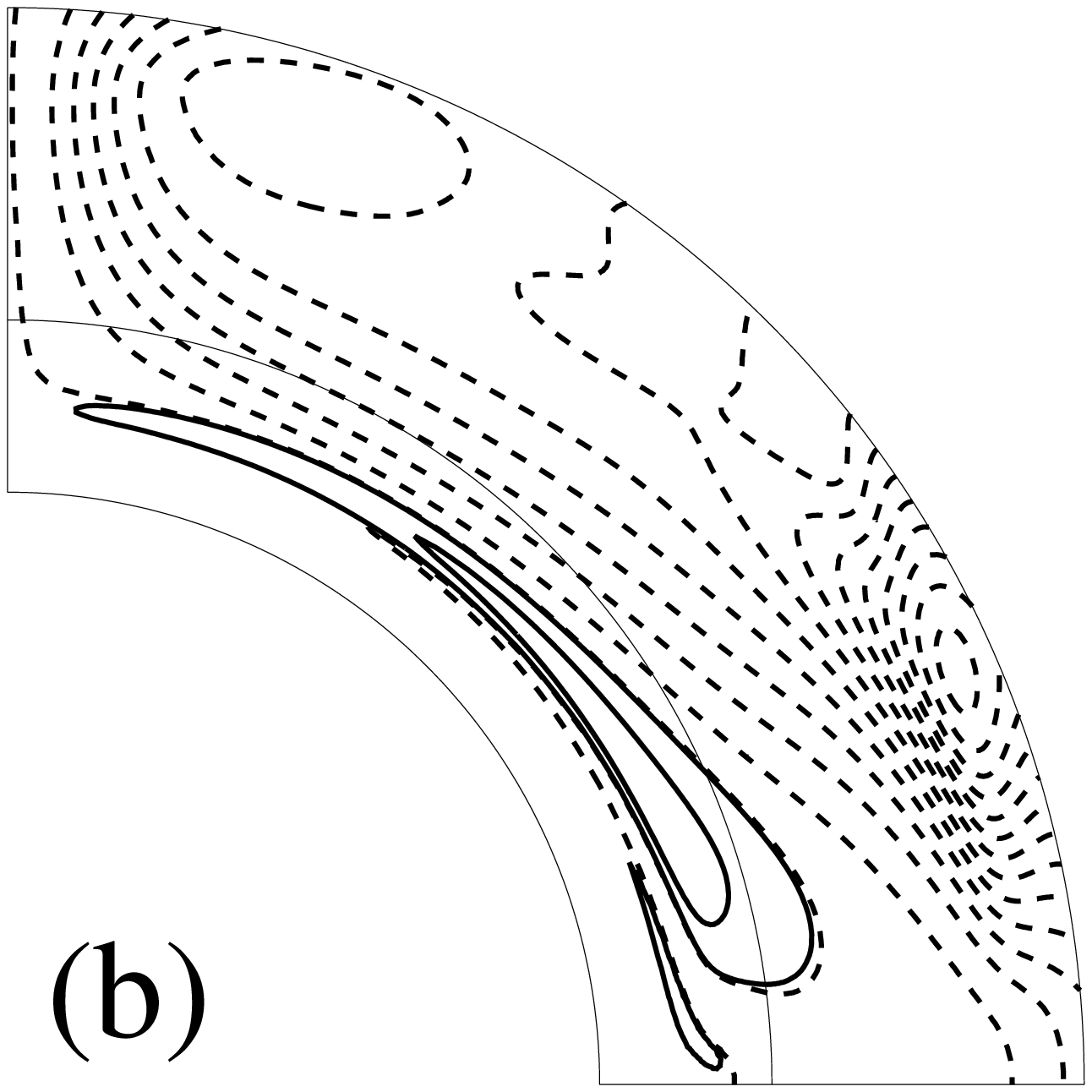}{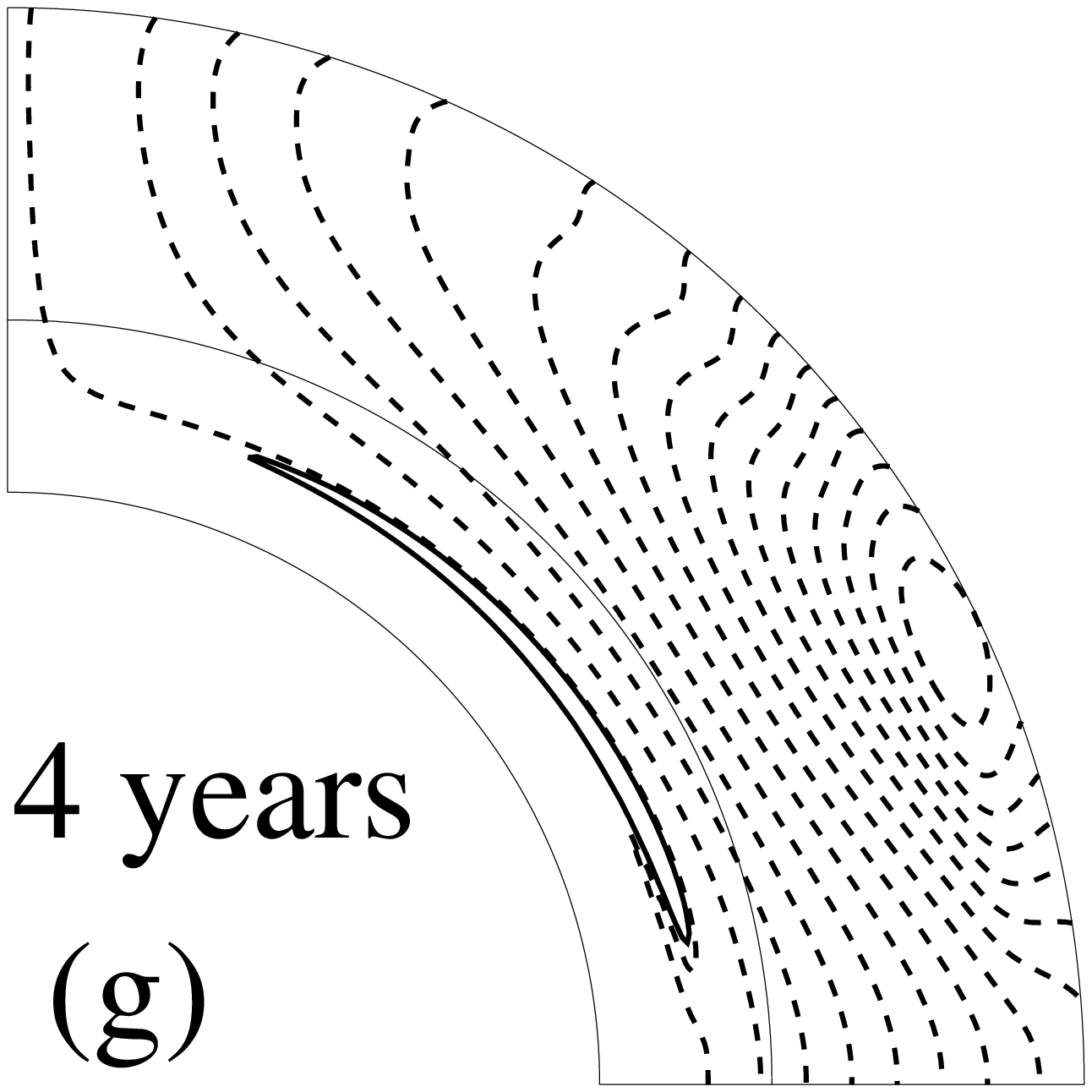}\\
\plottwo{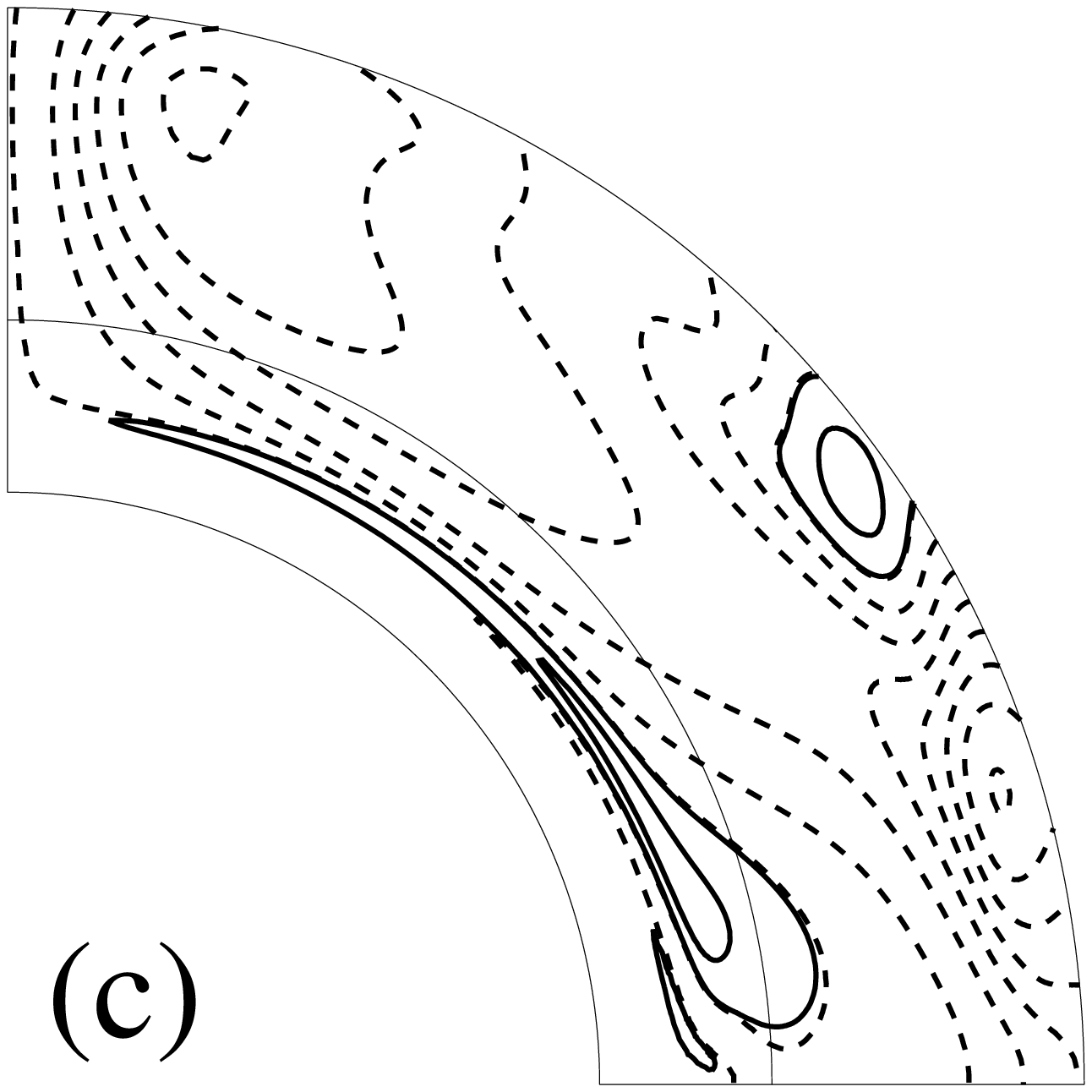}{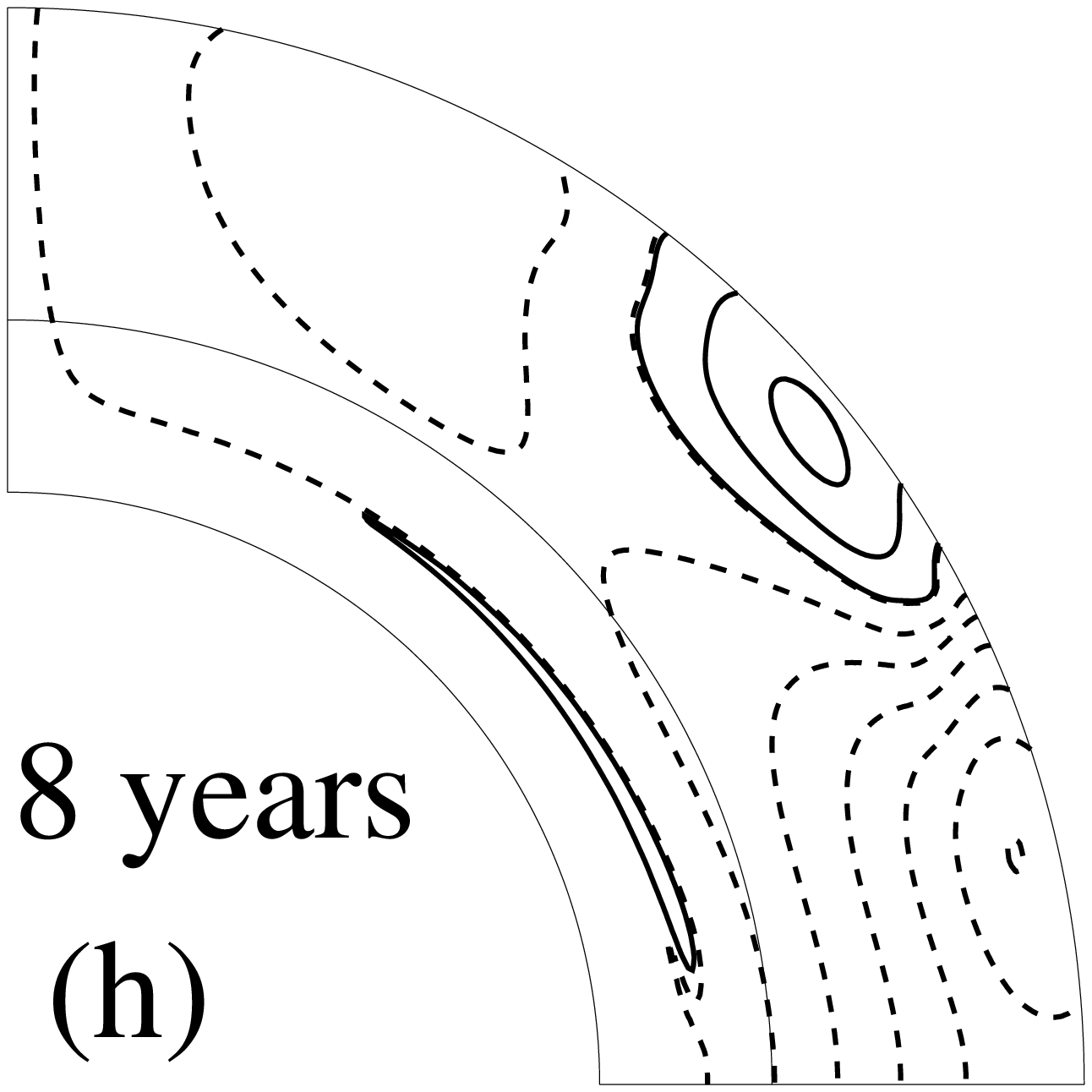}\\
\plottwo{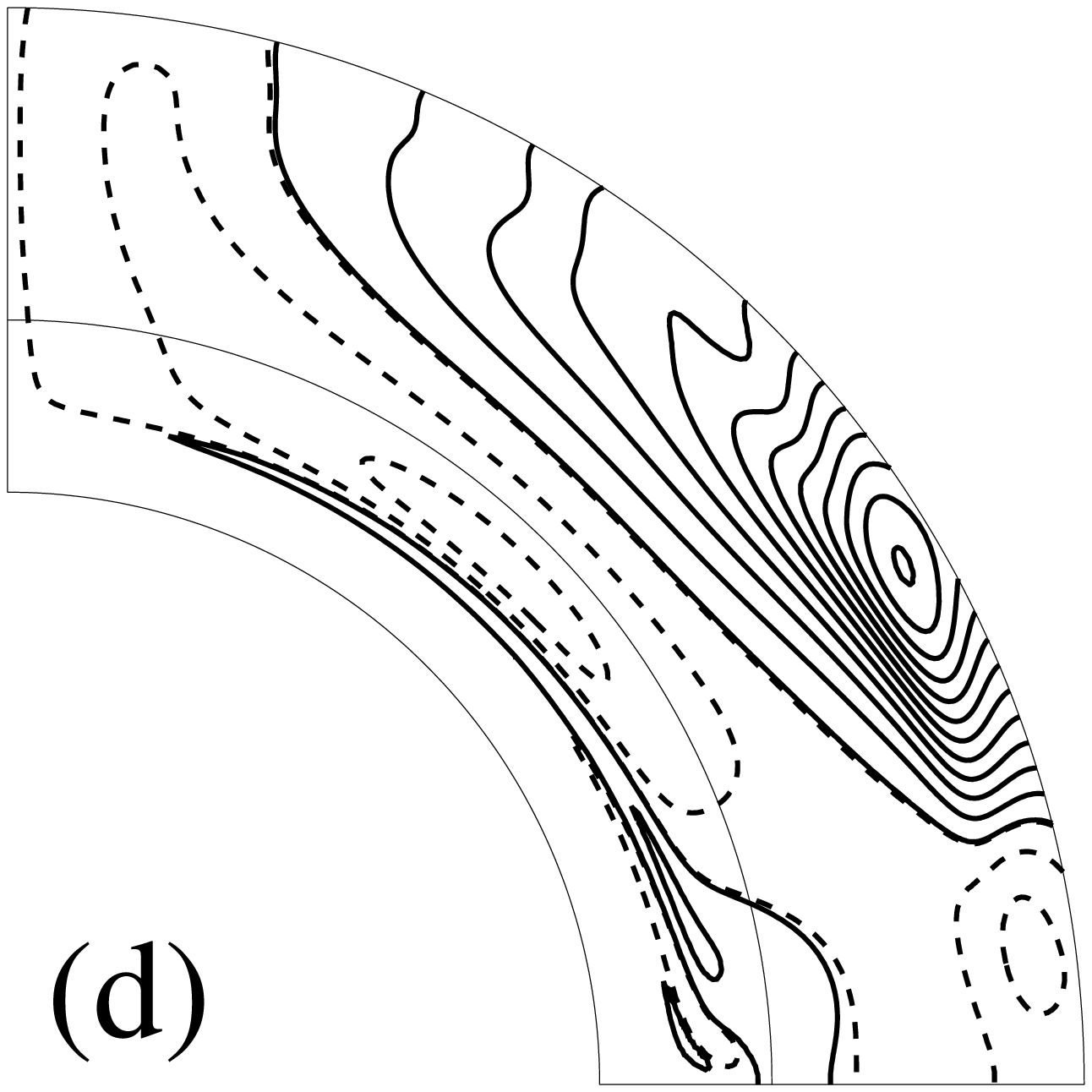}{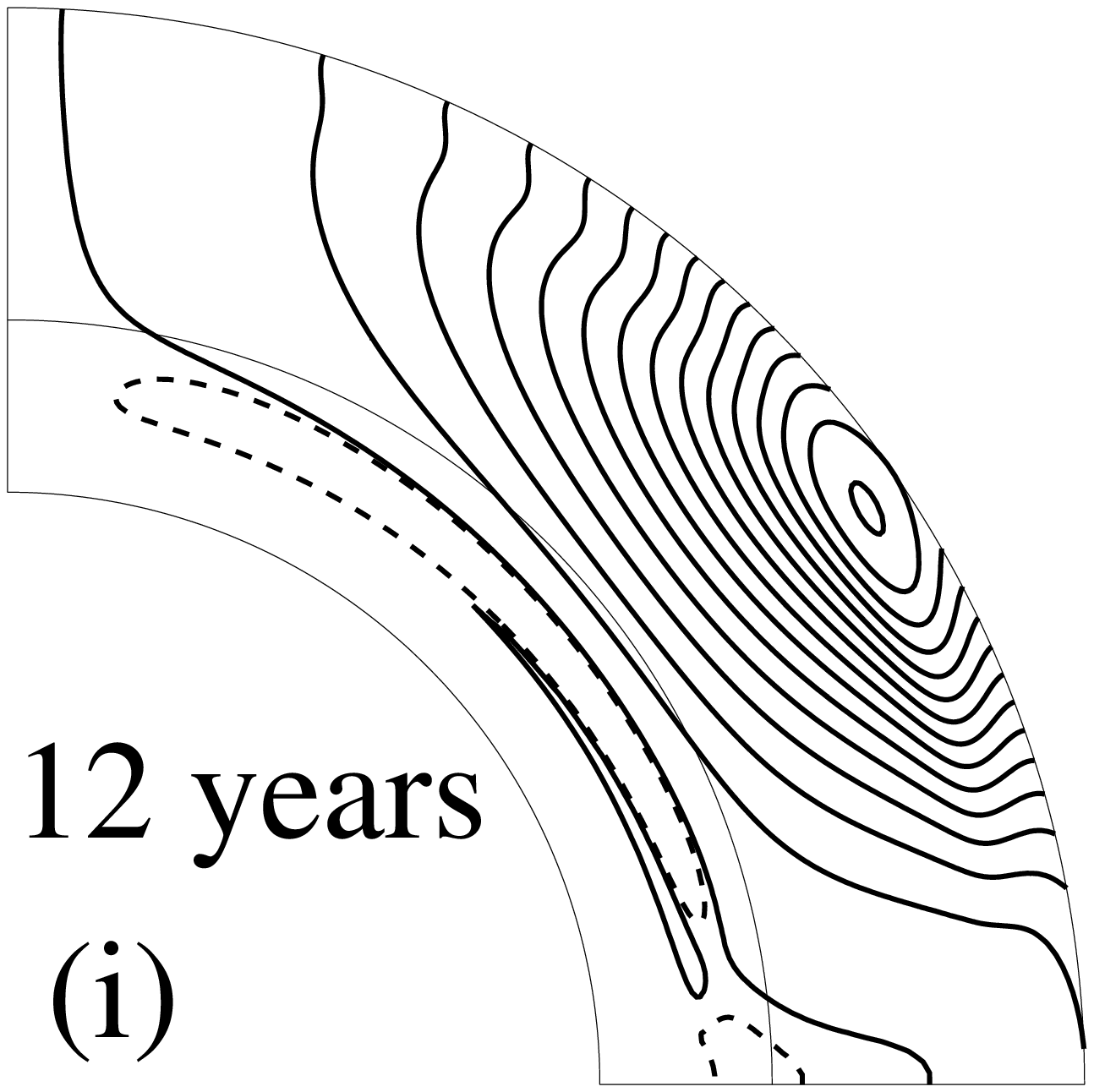}\\
\plottwo{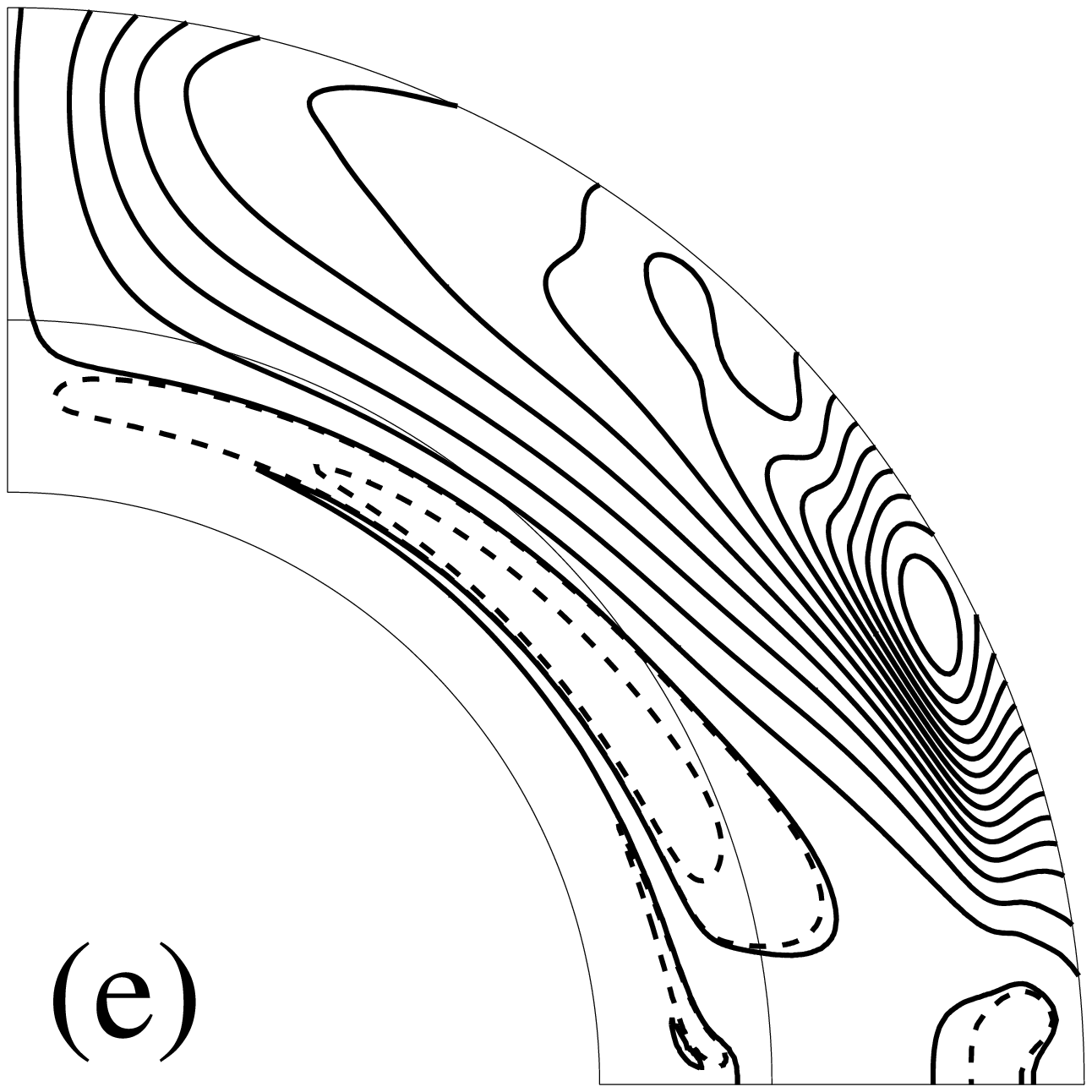}{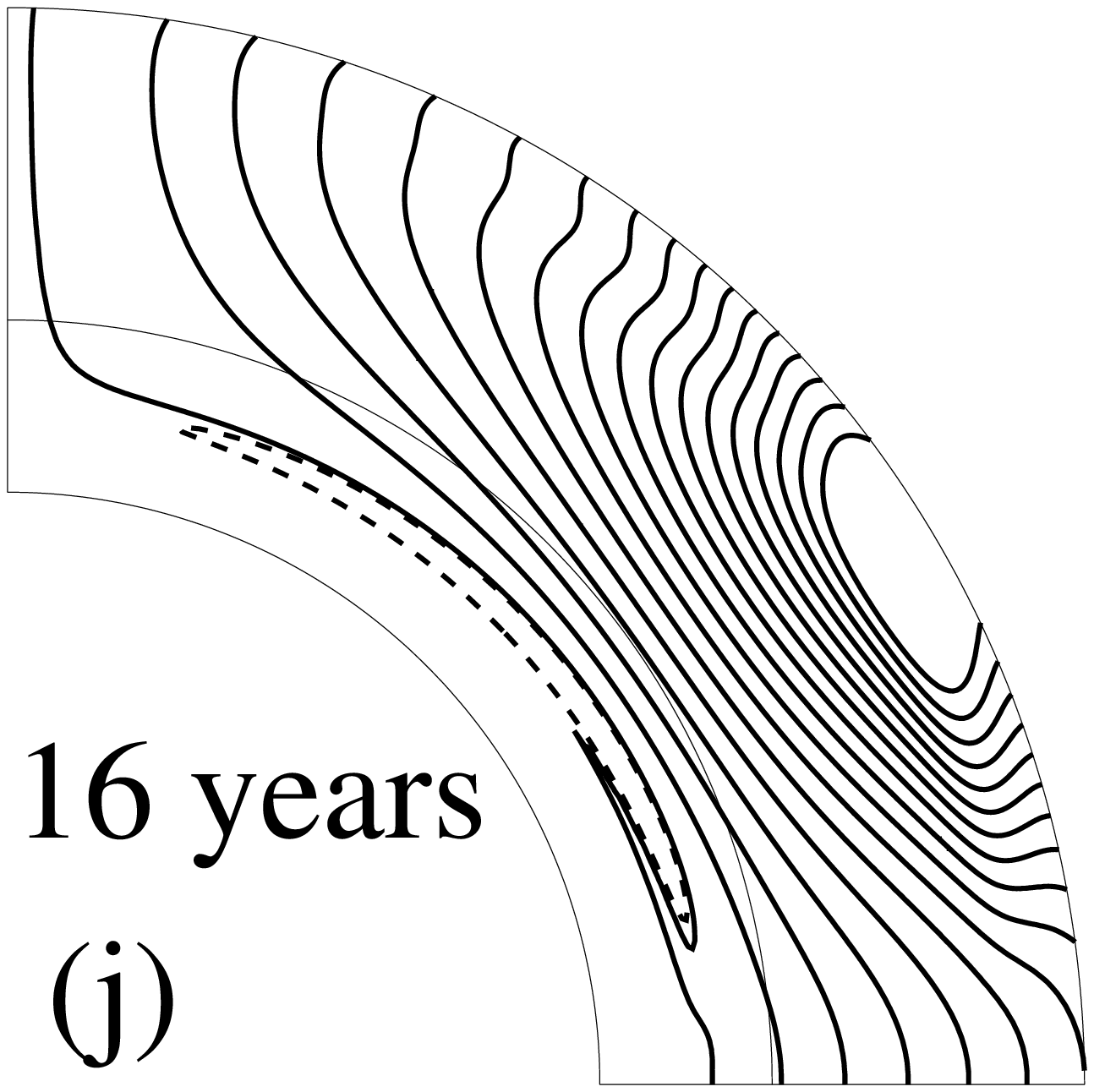}
\caption{Comparison
of poloidal fields in advection-dominated (left column/panels a to e) and
diffusion-dominated (right column/panels f to j) regimes. Each row corresponds to a
time through the solar cycle, running from one cycle minimum to the next. Solid lines
show clockwise field lines and dashed lines show anti-clockwise field lines. Also
indicated is the base of the SCZ at $0.71R_\odot$.} \label{fig:adpoloidal}
\end{center}
\end{figure}

\begin{figure}
\begin{center}
\epsscale{0.45}
\plottwo{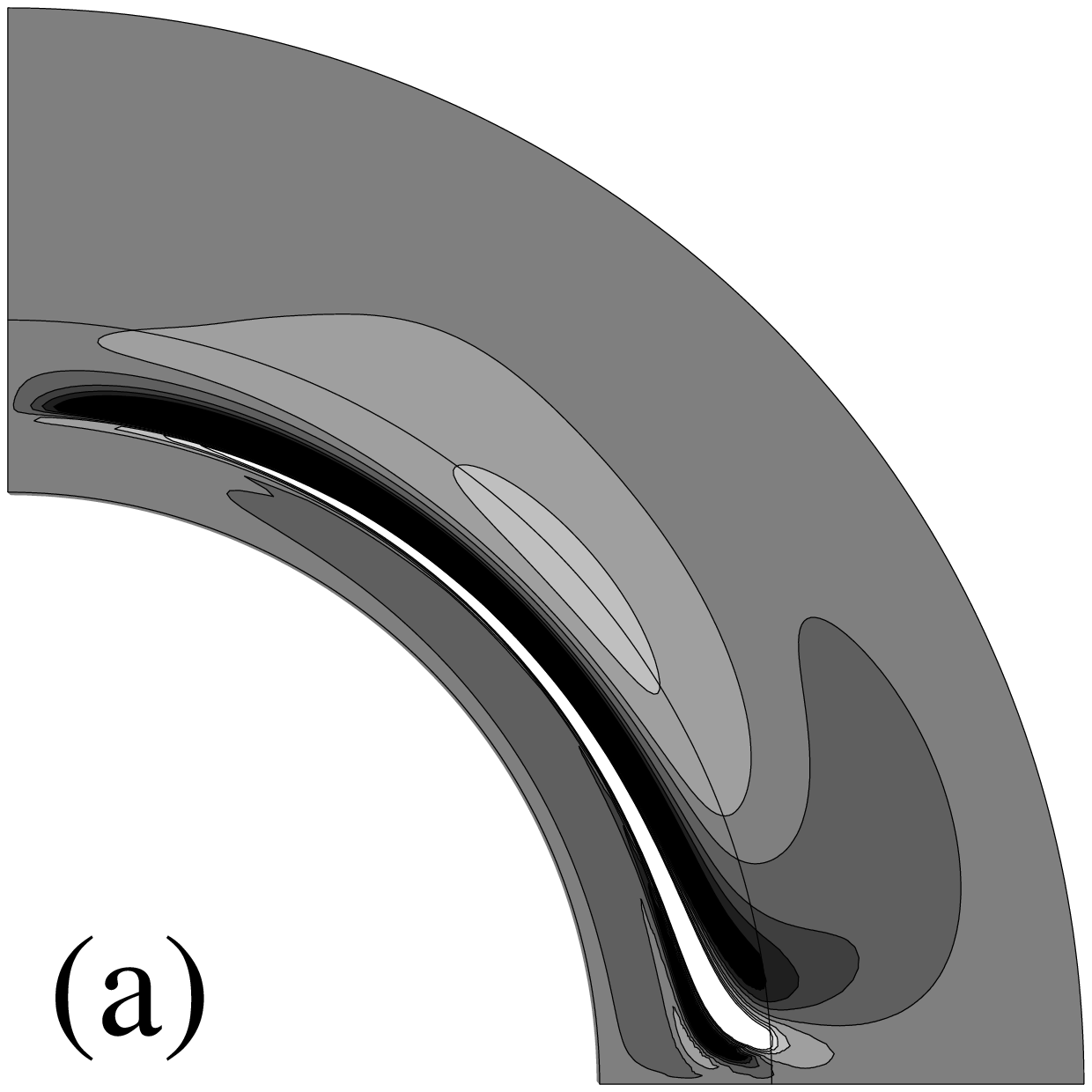}{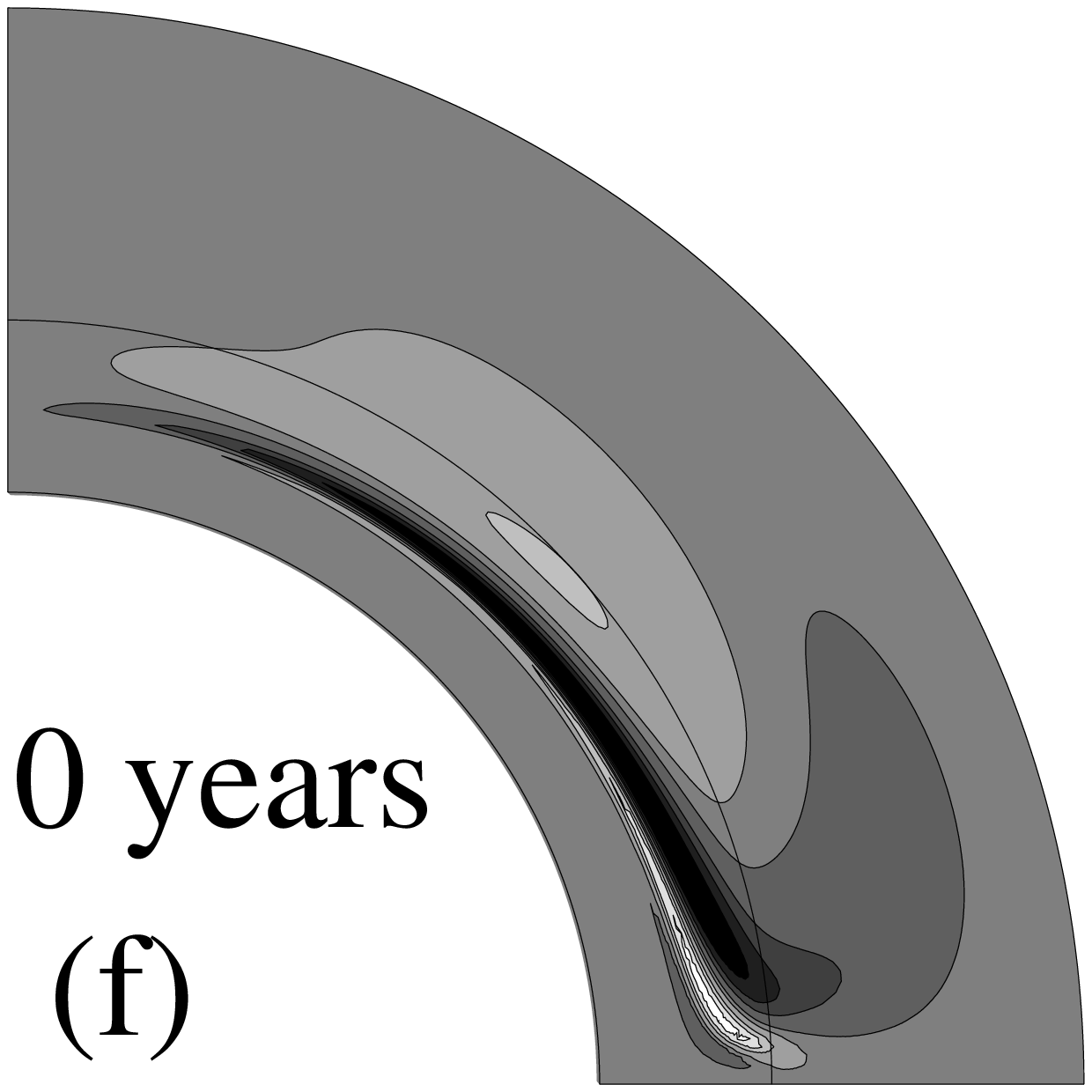}\\
\plottwo{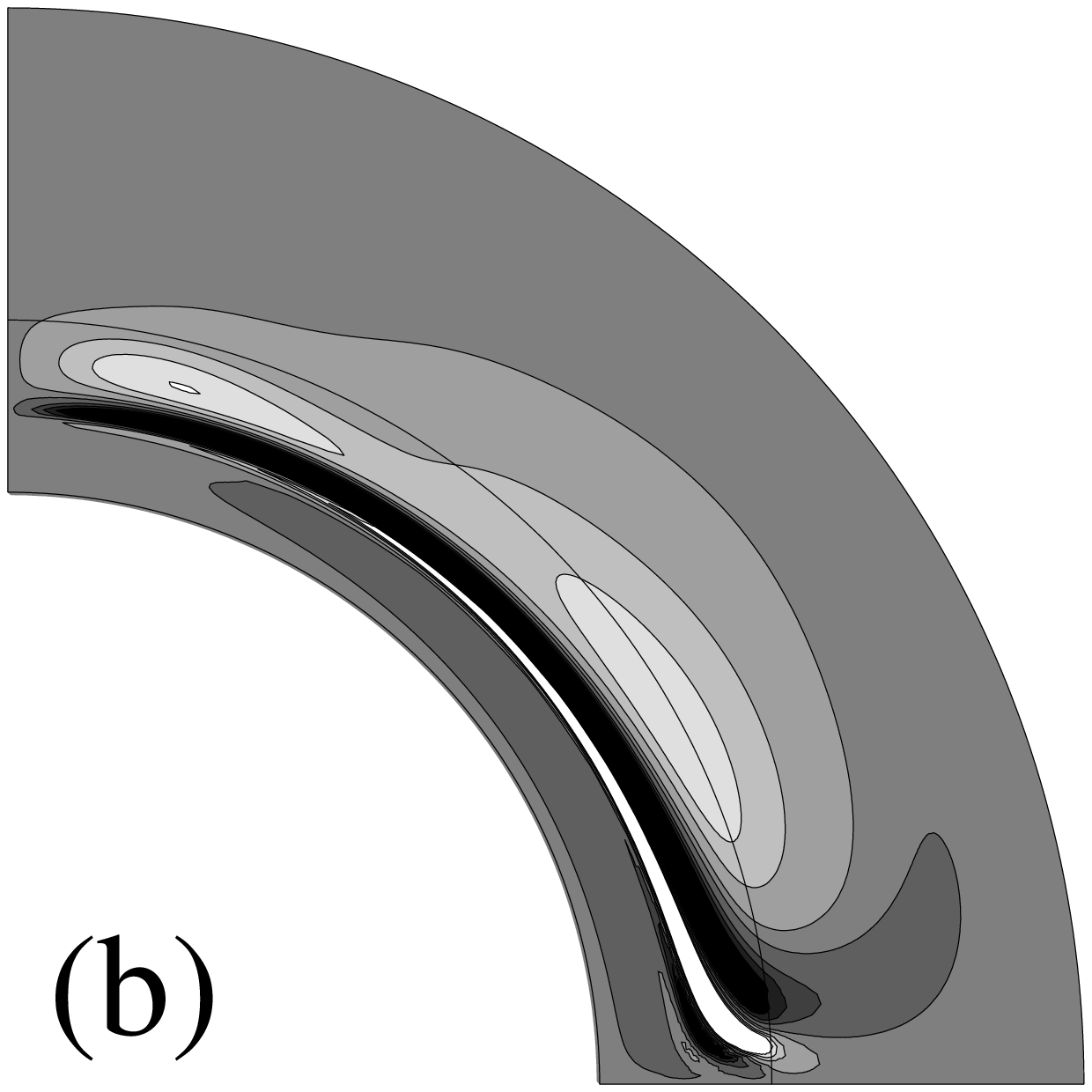}{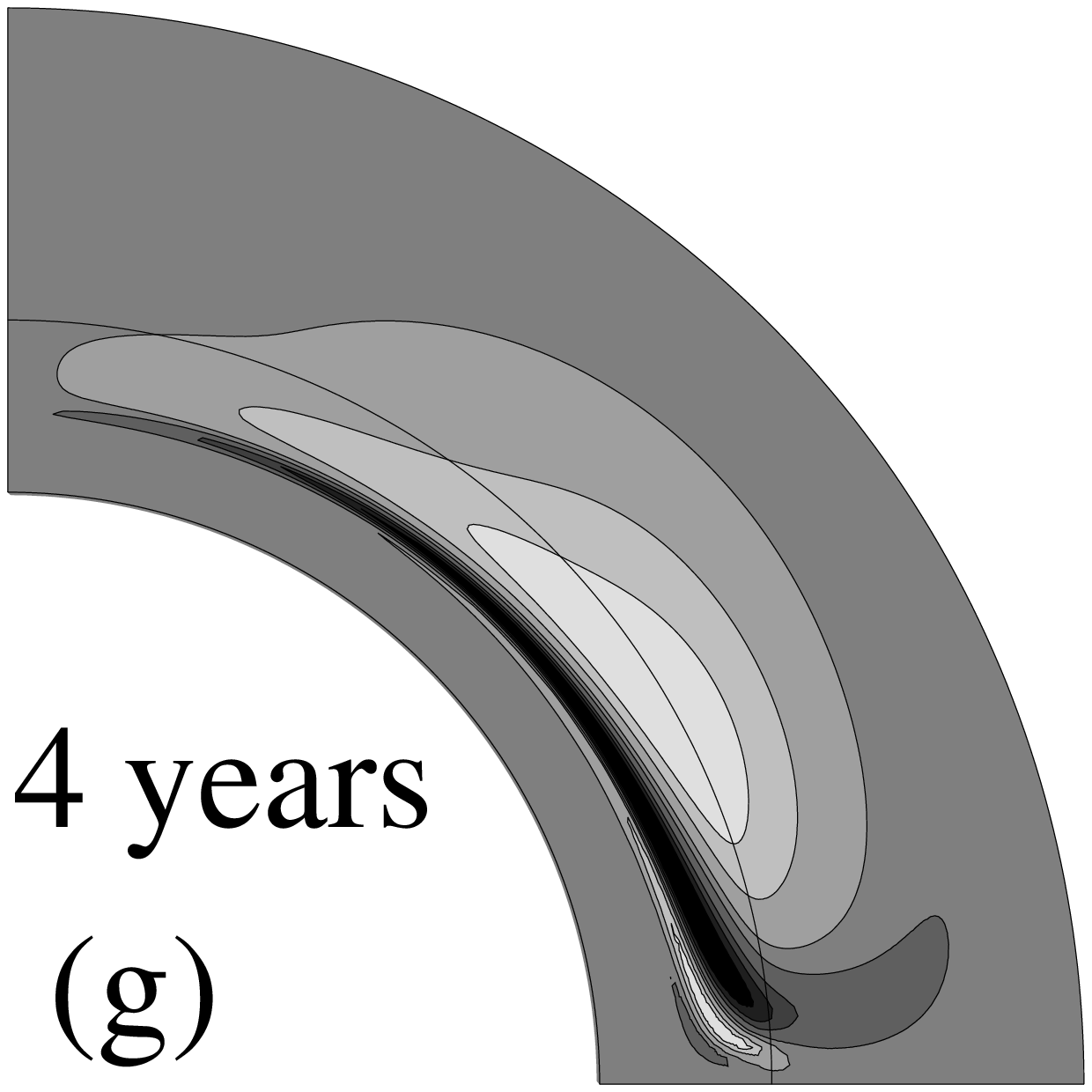}\\
\plottwo{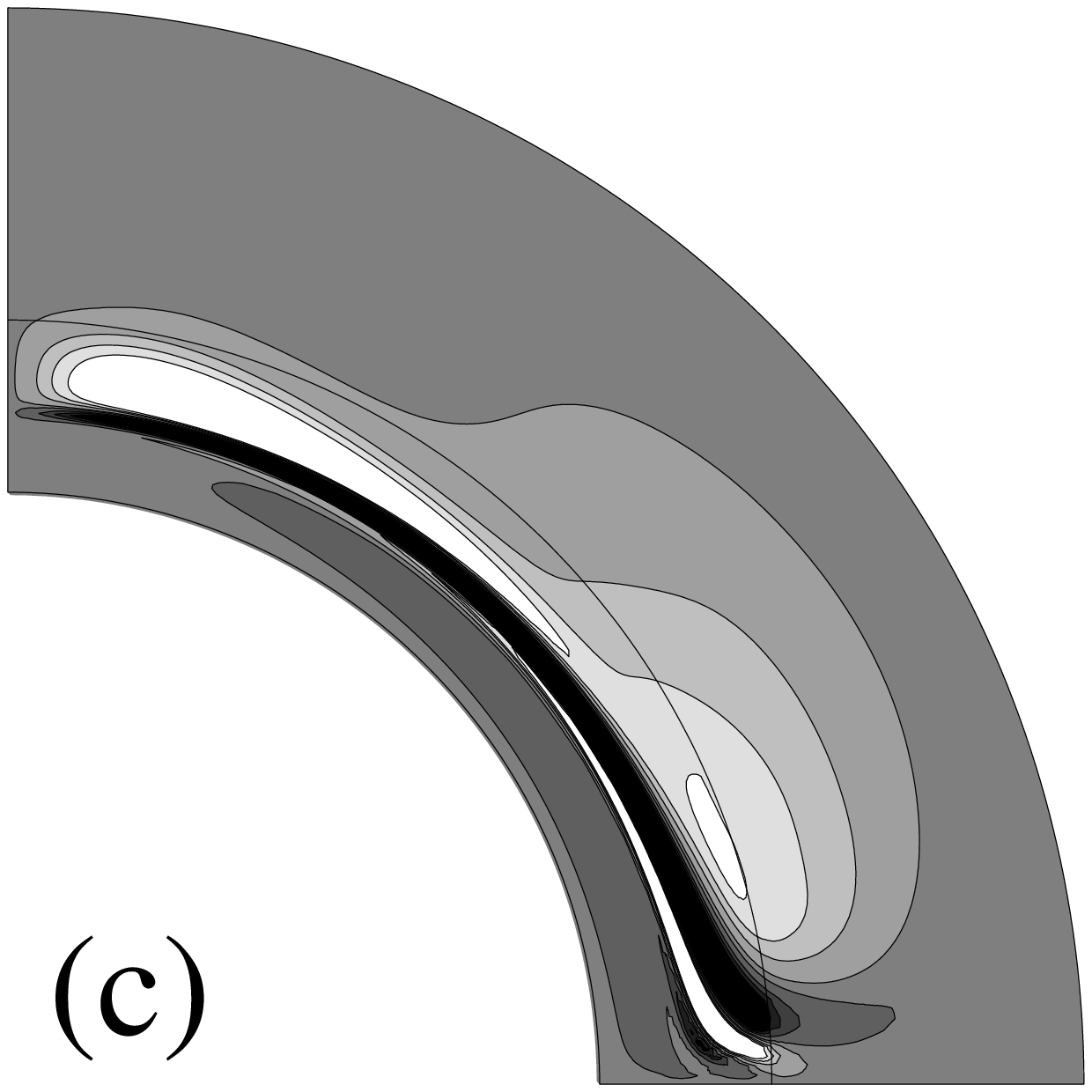}{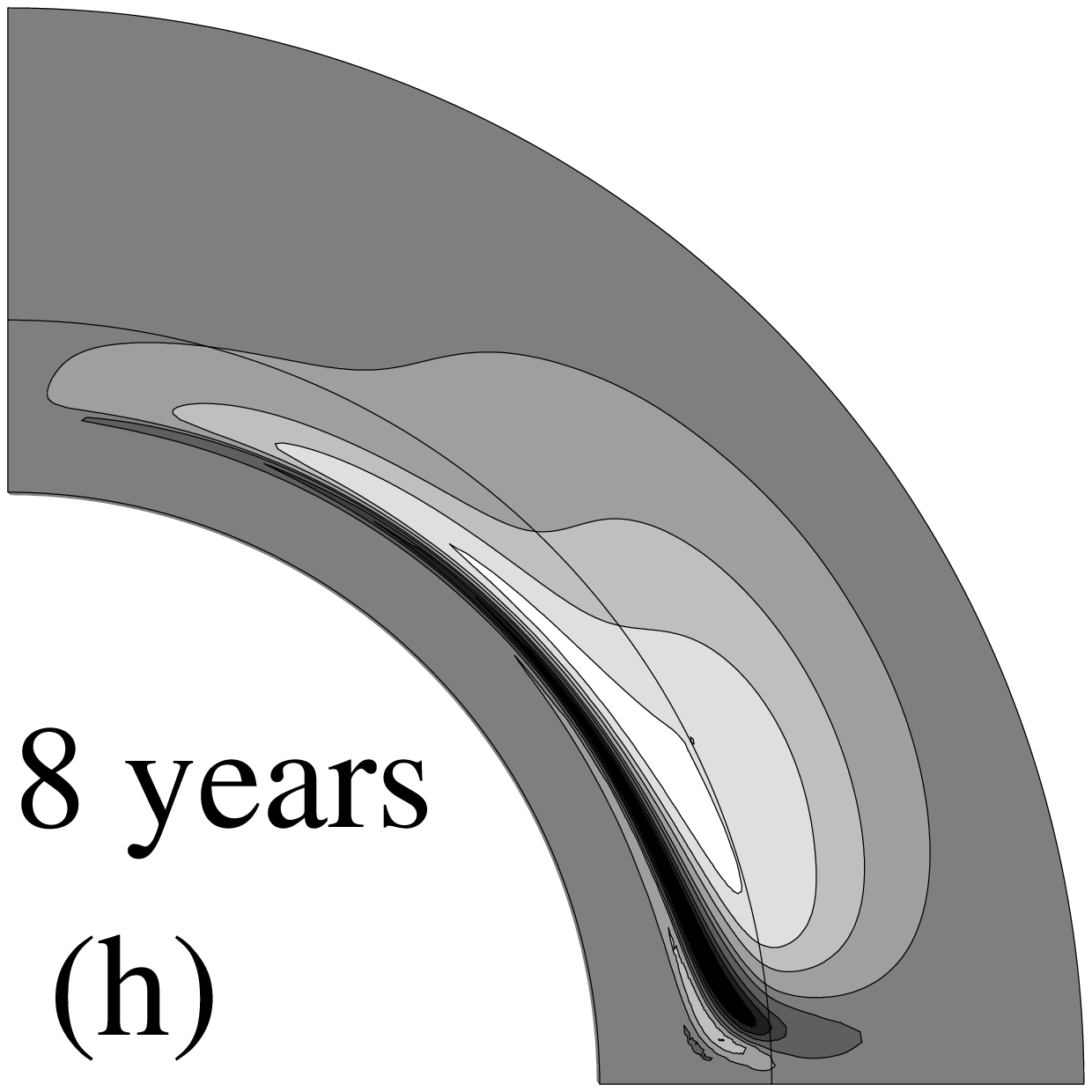}\\
\plottwo{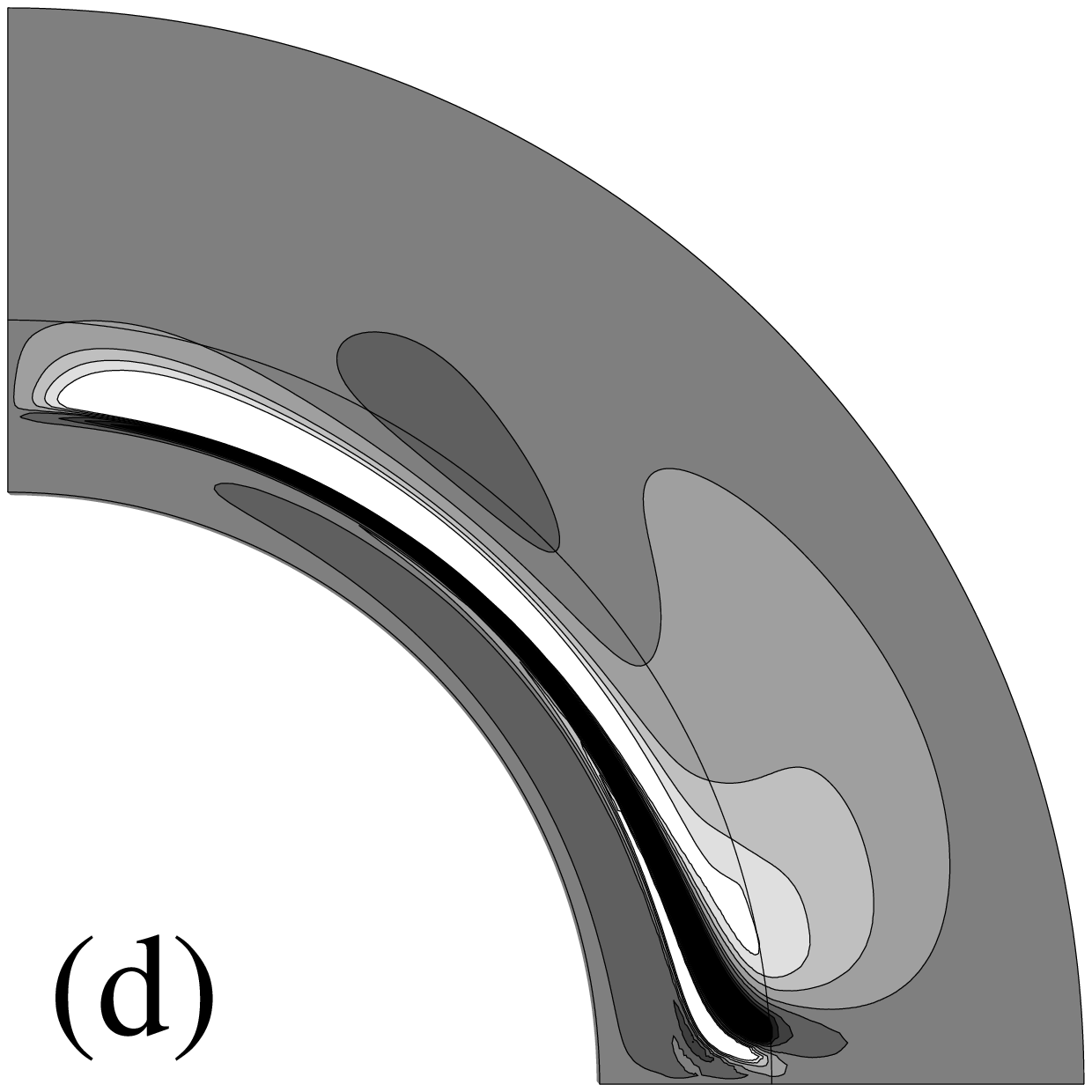}{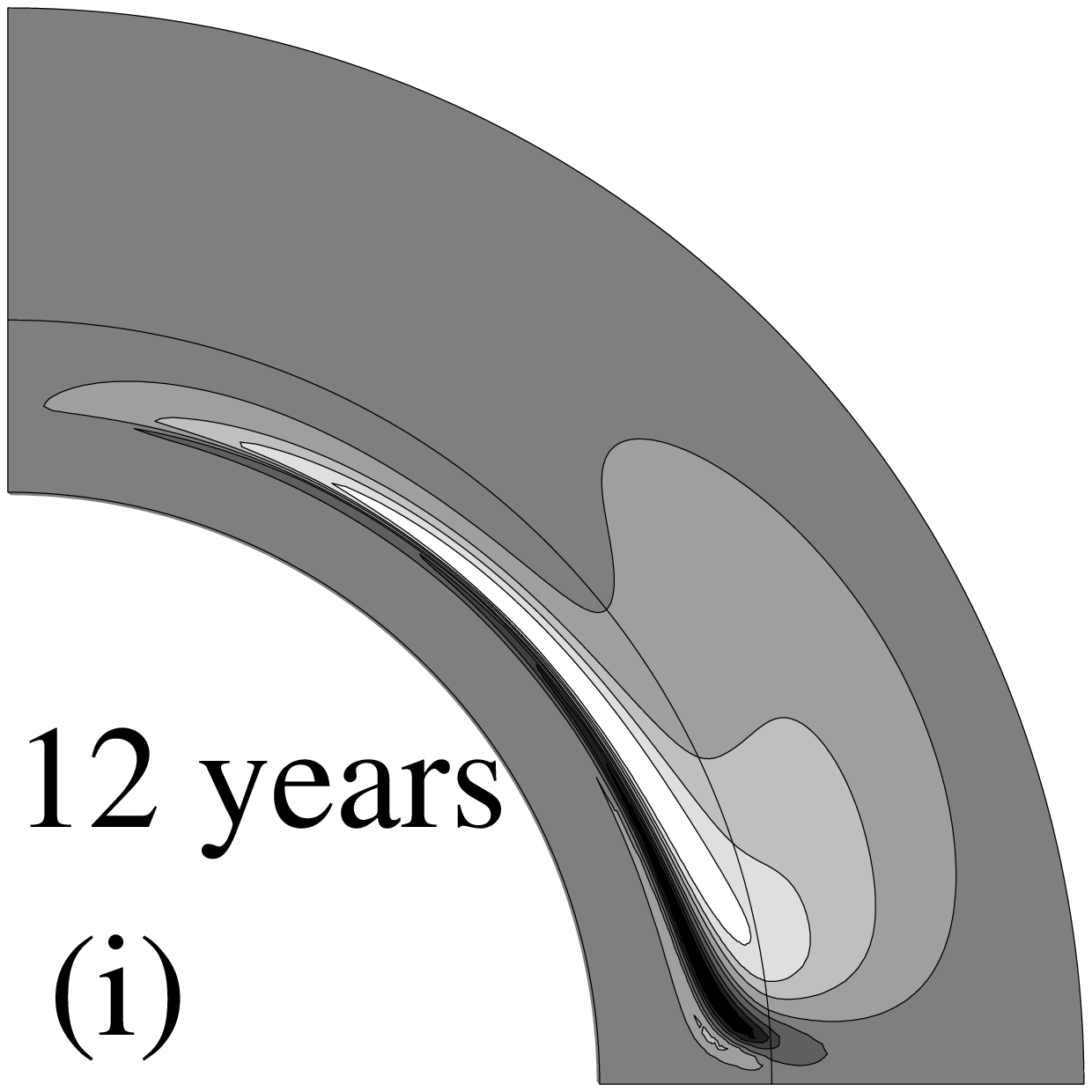}\\
\plottwo{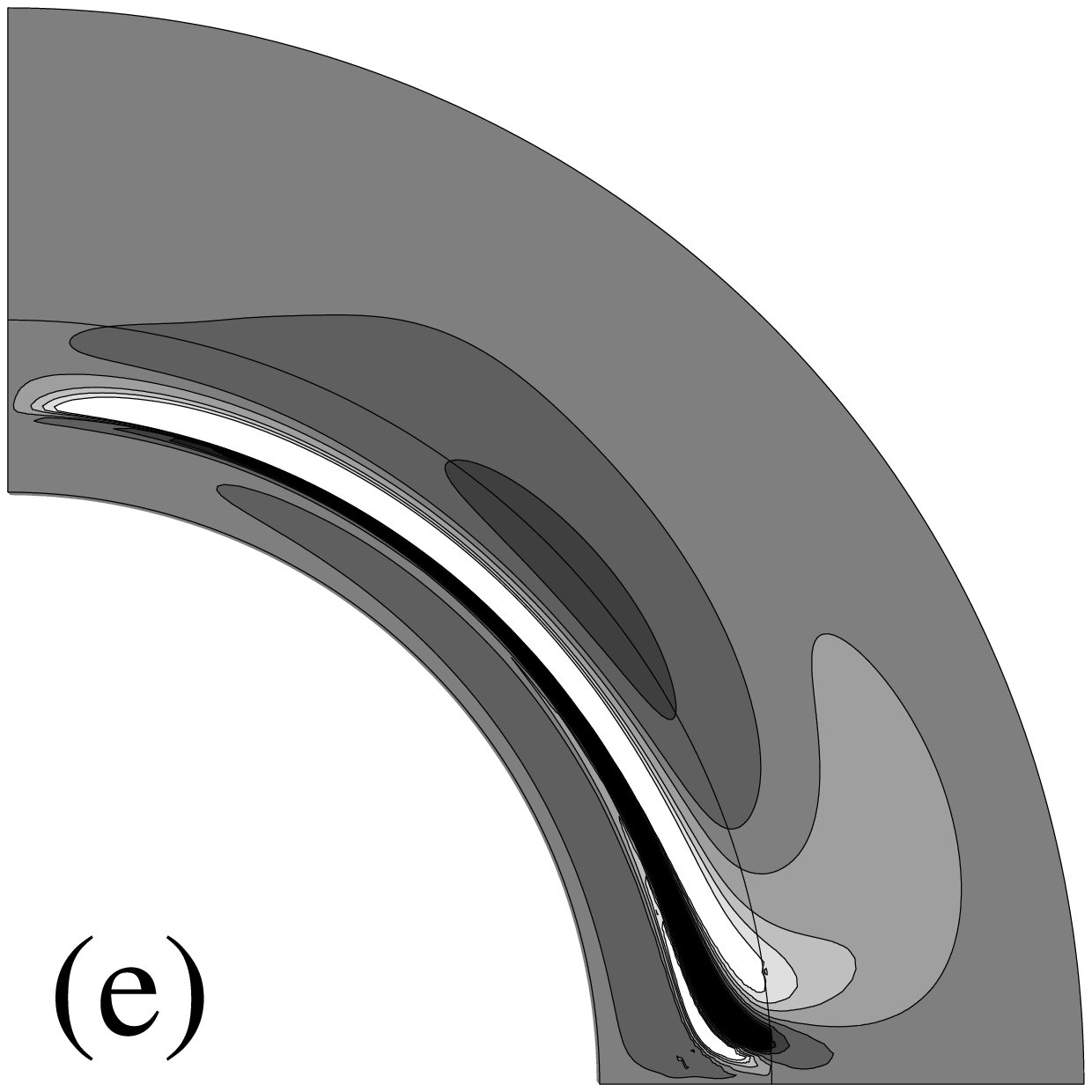}{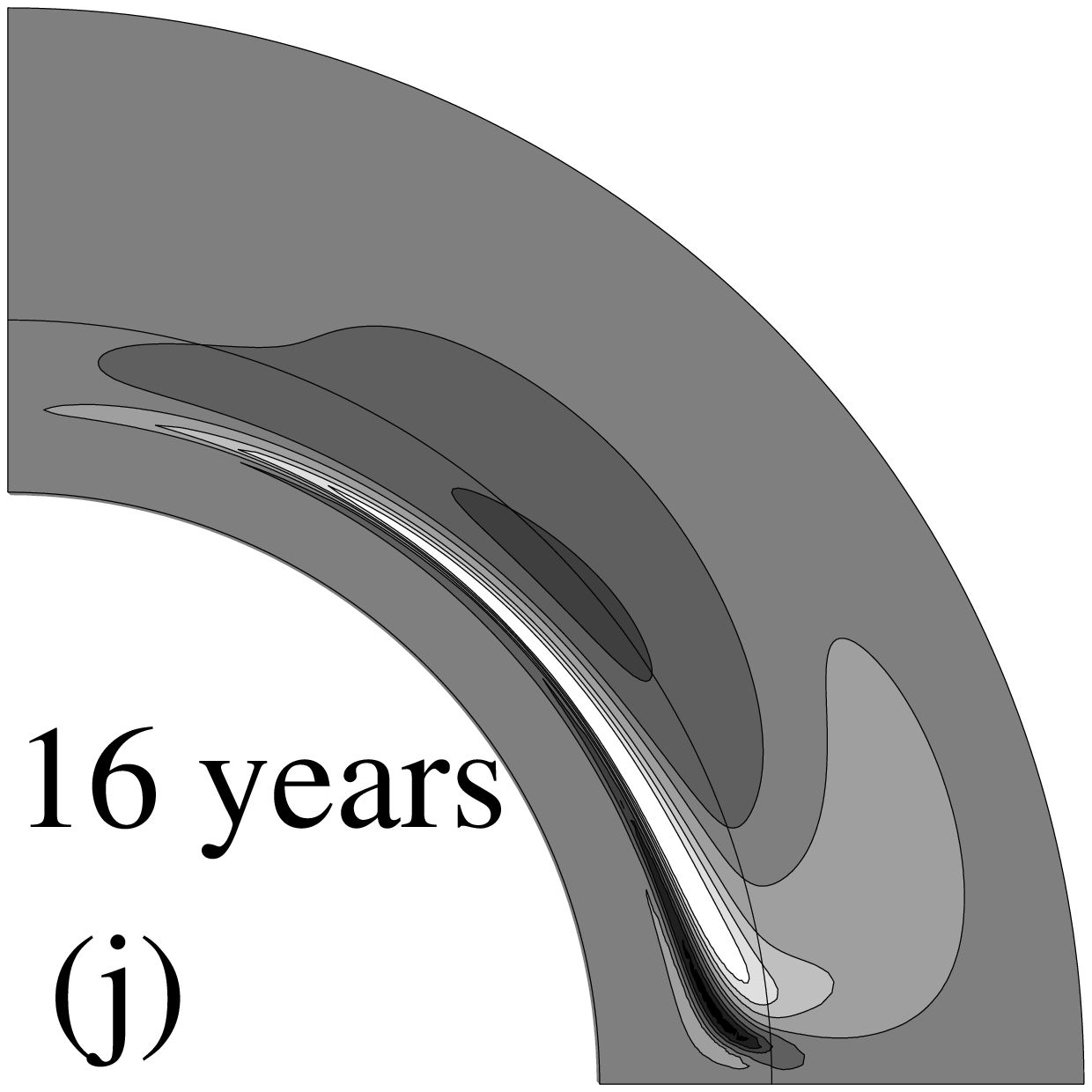}\\
\caption{Comparison
of toroidal field in advection-dominated (left column/panels a to e) and
diffusion-dominated (right column/panels f to j) regimes. Each row corresponds to a
different time through the solar cycle, running from one cycle minimum to the next.
Grayscale contours show toroidal field strength, with black corresponding to the
strongest negative field and white to the strongest positive toroidal field. Also
indicated is the base of the SCZ at $0.71R_\odot$.} \label{fig:adtoroidal}
\end{center}
\end{figure}

We now compare the poloidal and toroidal field evolution in the two regimes. Figure
\ref{fig:adpoloidal} shows the poloidal field lines for two runs, at different times
through the cycle, starting from one cycle minimum and finishing at the next cycle
minimum, so that the fields reverse in sign. The left-hand column is taken from a run
with $v_0=20\mpsec$ and $\eta_2=0.5\times 10^{12}\cmsqpsec$, which is in the
advection-dominated regime. The right-hand column is from a diffusion-dominated run
with the same $v_0$ but with $\eta_2=2.0\times 10^{12}\cmsqpsec$. Figure
\ref{fig:adtoroidal} shows the evolution of the toroidal field for the same two runs.


The key difference between the two regimes is the rate at which the poloidal field is
able to diffuse through the convection zone, after it is generated at the surface by
the Babcock-Leighton $\alpha$-effect. This is seen clearly by comparing the poloidal
field evolution between 8 and 12 years in the two runs (Figures \ref{fig:adpoloidal}c/h
and d/i). In the diffusion-dominated run the new clockwise poloidal field diffuses
directly down to the tachocline at all latitudes between these two times. However, in
the advection-dominated run the new poloidal field does not reach the tachocline until
the end of the cycle (16 years), and reaches the high latitudes before it reaches the
tachocline; i.e., in this case, the field evolution follows the meridional circulation
conveyor belt. There is still significant anticlockwise poloidal field remaining below
the tachocline from the previous cycle, and even a lower band of clockwise field from
the cycle before that. In the diffusion-dominated case there is only a weak band of
anticlockwise field remaining from the previous cycle at solar minimum.

This suggests that in the advection-dominated regime, poloidal fields from cycles $n$,
$n-1$, and $n-2$ combine to produce the toroidal field for cycle $n+1$, while in the
diffusion-dominated regime it is produced primarily from cycle $n$ poloidal field, with
a small contribution from cycle $n-1$.

The main difference in toroidal field evolution seen in Figure \ref{fig:adtoroidal} is
during the rising phase of the cycle, seen at 4 and 8 years (panels b/g and c/h). In
the advection-dominated regime there are two separate regions of toroidal field
generation, one region at high latitudes from poloidal field which has been advected
poleward by the meridional circulation, and a second region at lower latitudes arising
from direct diffusion of poloidal field across the convection zone. In the
diffusion-dominated case there is no strong generation of toroidal field at high
latitudes. In this case the strongest field generation occurs at mid to low latitudes,
with direct diffusion presumably being the primary means of transporting poloidal field
to the base of the convection zone (i.e., the meridional circulation conveyor belt is
``short-circuited''). We point our however that, contrary to usual expectations,
diffusive flux transport still plays a role in the advection-dominated case, and it is
responsible for the complex dependence of cycle amplitude on diffusivity in the
advection-dominated regime; we explore this issue below.

\subsection{The Role of Diffusive Flux Transport}

We have thus far identified the turnover in cycle amplitude to lie at the transition
point between advection-dominated and diffusion-dominated regimes of the dynamo. Within
the umbrella of this model, this maximum in the amplitude is understood to be a balance
between the time available for toroidal field amplification and the time available for
poloidal field decay. Figure \ref{fig:amp_diff} shows how the cycle amplitude varies
with the diffusivity $\eta_2$ for a fixed circulation speed $v_0$. We see that there is
a turnover in cycle amplitude for some value of $\eta_2$, with lower diffusivities
corresponding to the advection-dominated regime, and higher diffusivities to the
diffusion-dominated regime. In the diffusion-dominated regime, which is only reached
for lower speeds $v_0$ in Figure \ref{fig:amp_diff}, the cycle amplitude decreases with
increasing diffusivity. This is expected due to increased cancellation and decay of the
poloidal field. However, in the advection-dominated regime, the cycle amplitude
\emph{increases} as the diffusivity $\eta_2$ is increased. This initially seems
counter-intuitive, but we show here that it is caused by the influence of direct
diffusive flux transport of poloidal field across the convection zone.

\begin{figure}
\epsscale{0.8}
\plotone{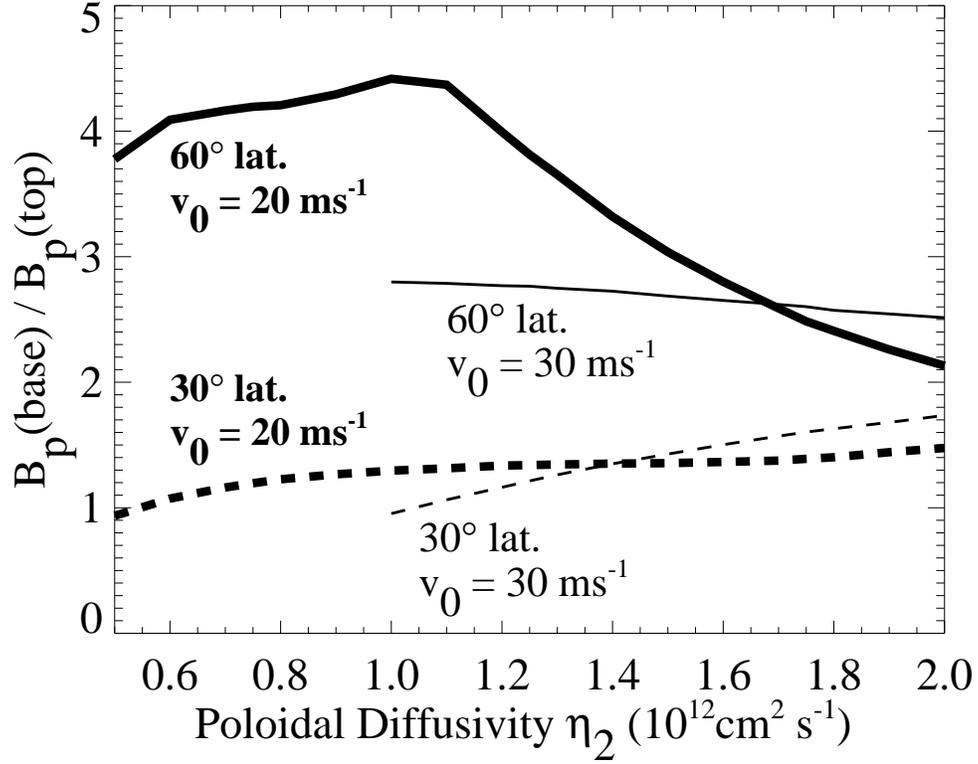}
\caption{Ratio of poloidal field $\left|\bpol\right|$ at base of convection zone
($r=0.715R_\odot$) to that at the surface ($r=R_\odot$), measured as a function of
diffusivity at latitudes $30^\circ$ (solid lines) and $60^\circ$ (dashed lines). Thick
lines correspond to runs with $v_0=20\mpsec$ and thin lines to runs with
$v_0=30\mpsec$. } \label{fig:polratio}
\end{figure}

This direct diffusion (from the solar surface to the base of the SCZ) was visible in
the poloidal field evolution plots shown in the previous section; we now demonstrate
its quantitative effect as $\eta_2$ is varied, by comparing the poloidal field strength
$\left|\bpol\right|$ at the base of the convection zone ($r=0.715R_\odot$) with that at
the solar surface ($r=R_\odot$). We take the ratio $\left|\bpol(\textrm{base})\right| /
\left| \bpol(\textrm{top})\right|$, using the peak value of $\left| \bpol\right|$ at
each location during the solar cycle. This ratio is plotted in Figure
\ref{fig:polratio}, measured at latitudes $30^\circ$ and $60^\circ$, and for two
different circulation speeds. Thin lines correspond to $v_0=30\mpsec$, where the dynamo
is in the advection-dominated regime for the whole range of $\eta_2$ shown. Thick lines
correspond to $v_0=20\mpsec$, for which the dynamo changes between the two regimes at
about $\eta_2=1.1\times 10^{12}\cmsqpsec$.


Consider first the behaviour at $30^\circ$ latitude (solid lines in Figure
\ref{fig:polratio}). Here the curves for both $v_0$ have positive slope, which implies
that a greater proportion of poloidal field from the surface reaches the bottom of the
convection zone as $\eta_2$ is increased. Thus direct diffusive flux transport at lower
latitudes always acts to increase the amount of poloidal field reaching the base of the
convection zone. Nearer to the pole, at $60^\circ$ latitude (dashed lines in Figure
\ref{fig:polratio}), the behaviour is different. Here the ratio decreases as $\eta_2$
is increased, both for the curve with $v_0=30\mpsec$ and in the diffusion-dominated
regime for $v_0=20\mpsec$. In the advection-dominated regime for $v_0=20\mpsec$
however, the ratio first increases with $\eta_2$. This suggests a more complex relation
between the surface and tachocline poloidal fields at high latitude, with competing
influence from both diffusive and advective flux transport. This is expected because
the downflow in the circulation is located at high latitudes.

The analysis presented in this section supports the idea that direct diffusive
transport of poloidal field across the convection zone, especially around mid-latitudes, is responsible for the trend of increasing cycle amplitude with increasing
diffusivity, found in the advection-dominated regime. Although such diffusive transport
acts to increase cycle amplitude in both regimes, diffusion also causes the poloidal
field that is being transported by meridional circulation to decay, cancelling with
field from the previous cycle that is stored below the tachocline. Thus diffusion also
has a negative effect on cycle amplitude. It is this negative effect which dominates at
higher diffusivities, forcing the dynamo into the diffusion-dominated regime where
cycle amplitude decreases with increasing diffusivity.

\section{Persistence of Memory: Cycle-to-Cycle Correlations in Advection versus
Diffusion Dominated Regimes in a Stochastically Forced Dynamo} \label{sec:timedelays}

It is expected that the memory of a flux-transport dynamo is much longer in the
advection-dominated regime than in the diffusion-dominated regime, and solar cycle
predictions have been based on this expectation \citep{dikpati2006b,jiang2007}.
However, a detailed comparative analysis of persistence of memory in these different
regimes under the umbrella of the same model had not been previously performed. Our
analysis in the previous section has brought us closer to understanding the flux
transport dynamics (in these two regimes) that is the physical basis for any memory
mechanism. In this section, we consider how the persistence of this memory differs
between the two regimes, by looking at the correlation between peak polar and toroidal
fields of subsequent cycles. Since the simulations considered earlier relaxed to a
regular periodic cycle, we cannot use these to study correlations between different
cycles. Therefore, we now introduce self-consistent fluctuations in the cycle
properties by means of a stochastically varying $\alpha$-effect, and explore the
resulting correlations between different cycles.

\subsection{Stochastic Fluctuations}
We introduce fluctuations in the model by varying the coefficient $\alpha_0$ of the
$\alpha$-effect (see \citealp{chatterjee2004} for the full expression). We set
\begin{equation}
\alpha_{0} = \alpha_{\textrm{base}} + \alpha_{\textrm{fluc}}\,\sigma(t; \tau_{\textrm{cor}}),
\end{equation}
where $\alpha_{\textrm{base}}=30\mpsec$ is the mean value,
$\alpha_{\textrm{fluc}}=30\mpsec$ gives the maximum amplitude of the fluctuations
(corresponding to the 200\% level), and $\sigma$ is a uniform random deviate selected
from the interval $[-1,1]$, with a new value after each coherence time
$\tau_{\textrm{cor}}$. Although for our purposes this is essentially a device for
changing the cycle properties from one cycle to the next, there is a strong physical
basis for stochastic variations in $\alpha$ which have been invoked in several previous
studies \citep{choudhuri1992,hoyng1993,ossendrijver1996,charbonneau2000,mininni2002}.
Our model uses a Babcock-Leighton $\alpha$-effect where poloidal field is generated at
the surface from the decay of tilted active regions \citep{babcock1961,leighton1969}.
Thus stochastic variations in the $\alpha$ coefficient are natural, because it arises
from the cumulative effect of a finite number of discrete flux emergence events (active
region eruptions with varying degrees of tilt).

To compare the two regimes we consider two runs, both with $\eta_{2}=1.0\times
10^{12}\cmsqpsec$. The circulation speed $v_0$ is kept constant throughout each run,
and only the $\alpha$ effect is varied. Run 1 has $v_0=15\mpsec$, so is
diffusion-dominated, while run 2 has $v_0=26\mpsec$ and is advection-dominated. The
coherence time $\tau_{\textrm{cor}}$ is set to $2.3$ years in run 1 and $1.5$ years in
run 2, so as to keep the ratio of the former to the cycle duration roughly the same in
each case. We note that although the exact value of the coherence time is not important
for our study (and is introduced just as a means to enable sufficient fluctuations),
the timescale -- on the order of a year -- is chosen to reflect that the BL $\alpha$-effect
is a result of surface flux transport processes (diffusion, meridional circulation and
differential rotation) which can take up to a year to generate a net radial (component
of the poloidal) field from multiple flux emergence events \citep{mackay2004}.

\subsection{Correlation Analysis}
In this section we compare the peak surface radial flux $\Phi_r$ for cycle $n$ with the
peak toroidal flux $\Phi_{\textrm{tor}}$ for cycles $n$, $n+1$, $n+2$, and $n+3$. The
toroidal flux is defined as before by integrating $B(r,\theta)$ over the region
$r=0.677R_\odot$ to $0.726R_\odot$, $\theta=45^\circ$ to $80^\circ$. The radial flux
$\Phi_r$ is found by integrating $B_r(R_\odot,\theta)$ over the solar surface between
$\theta=1^\circ$ to $20^\circ$, (i.e., latitudes $70^\circ$ to $89^\circ$). Note that
the peak toroidal flux precedes the peak surface radial flux for the same cycle, which
has the same sign. The poloidal field then produces the toroidal field for cycle $n+1$
with the opposite sign. We measure the correlation of the surface radial flux for cycle
$n$ with the toroidal flux of different cycles, comparing the absolute value of each
total signed flux.

\begin{figure}
\plotone{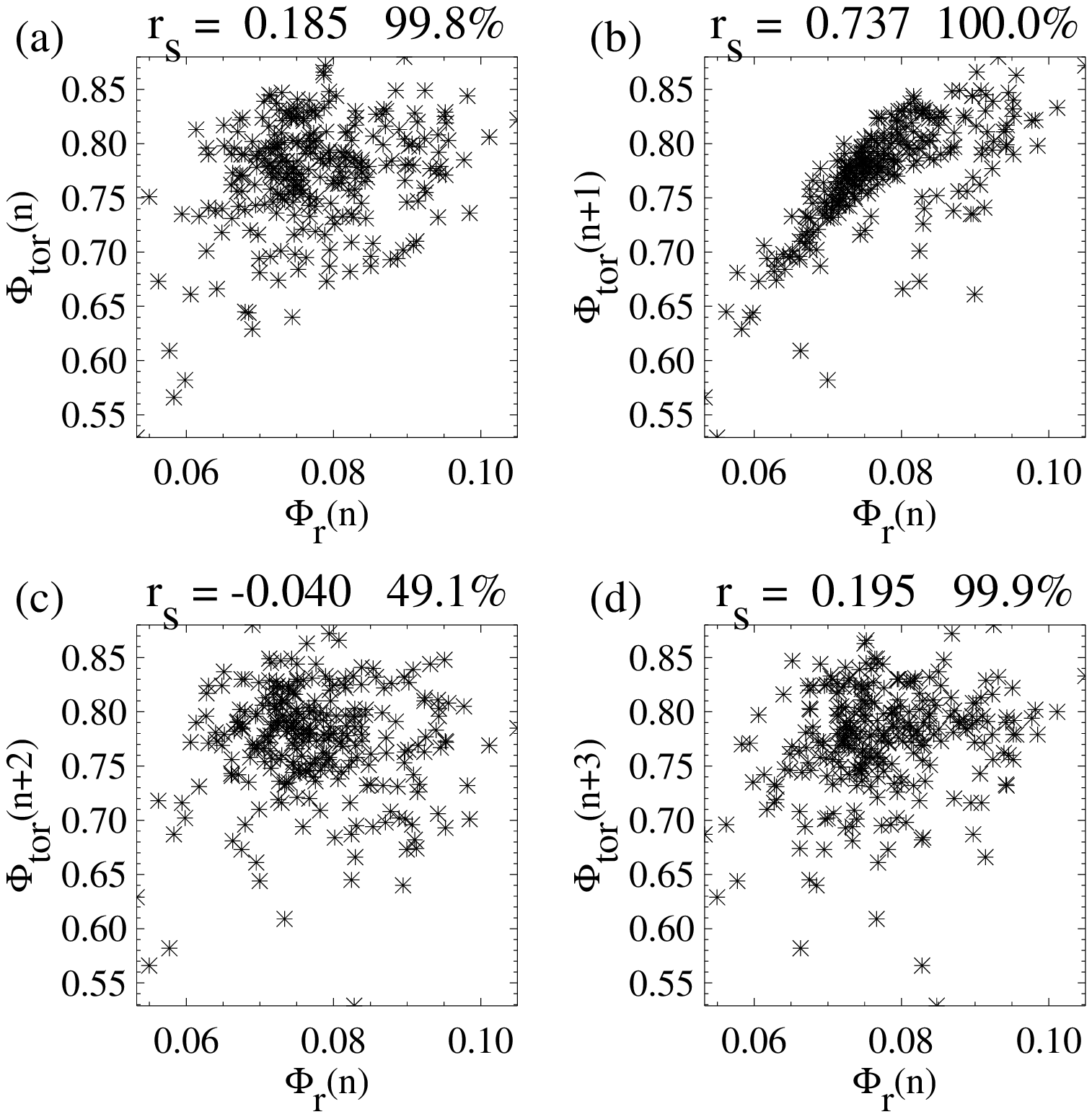}
\caption{Cycle-to-cycle correlations in the diffusion-dominated regime (run 1), between radial flux
$\Phi_r(n)$ and (a) toroidal flux $\Phi_{\textrm{tor}}(n)$, (b) $\Phi_{\textrm{tor}}(n+1)$,
(c) $\Phi_{\textrm{tor}}(n+2)$, and (d) $\Phi_{\textrm{tor}}(n+3)$. The Spearman's rank
correlation coefficient is given along with its significance level for 275 cycles. All
magnetic fluxes are in units of $10^{25}\mx$.} \label{fig:cordiff}
\end{figure}

Both runs were computed for a total of 275 cycles with fluctuating $\alpha_0$, so as to
produce meaningful statistics for each of the dynamo regimes. The results are
illustrated in Figures \ref{fig:cordiff} and \ref{fig:coradv} as scatter-plots of
$\Phi_\textrm{tor}$ for different cycles against $\Phi_r(n)$. The (non-parametric)
Spearman's rank correlation coefficient $r_\textrm{s}$ is given above each plot, along
with its significance level. The correlation coefficients are summarised in Table
\ref{tab:correlations}, where the Pearson's linear correlation coefficient
$r_\textrm{p}$ is also given for comparison. Although the latter is less reliable, as
it assumes a linear relation, it agrees well with $r_\textrm{s}$ in each case.



The results show a clear difference between the two regimes. The advection-dominated
regime shows significant correlations at all 4 time delays, apparently suggesting that
the memory of past poloidal field survives for at least 3 cycles; however, more on this
later. The diffusion-dominated regime has a strong correlation only between $\Phi_{r}(n)$
and $\Phi_{\textrm{tor}}(n+1)$, suggesting that the dominant memory relates to just a one
cycle time-lag, although very weak correlations are also found with
$\Phi_{\textrm{tor}}(n)$ and $\Phi_{\textrm{tor}}(n+3)$.

In both regimes the strongest relation is the positive correlation between $\Phi_{r}(n)$
and $\Phi_{\textrm{tor}}(n+1)$. This is to be expected as this is the more deterministic
phase of the cycle -- where toroidal fields (of cycle $n+1$, say) are inducted from the
older cycle $n$ poloidal field via the relatively steady differential rotation. Note
however that the two fluxes do not have to be directly coupled, in that the two fluxes
may be positively correlated because they are both created from the mid-latitude
poloidal field of cycle $n$ (generated by the $\alpha$-effect). The polar flux
$\Phi_{r}(n)$ arises through poleward meridional transport of the cycle $n$ poloidal
field, while the toroidal flux $\Phi_{\textrm{tor}}(n+1)$ is generated from cycle $n$
poloidal field that is diffusively transported across the convection zone. This is
particularly the case in the diffusion-dominated regime. Nonetheless, even this
indirect scenario suggests that the strongest correlation should in fact be between
the cycle $n$ poloidal field and cycle $n+1$ toroidal field in this class of
$\alpha$-$\Omega$ dynamo models.

The other phase of the cycle, in which the poloidal field is generated by the
$\alpha$-effect, is inherently more random due to the fluctuating $\alpha$-effect in
these runs. Nevertheless, there is a strong positive correlation between
$\Phi_{\textrm{tor}}(n)$ and $\Phi_{r}(n)$ in the advection-dominated regime, while this
correlation is largely absent in the diffusion-dominated regime. This we attribute to
the relatively stronger role of advective flux transport in the advection-dominated
regime -- which implies that a larger fraction of the original toroidal flux that has
buoyantly erupted is transported to the polar regions by the circulation. In effect
therefore, the advection-dominated regime allows correlations to propagate \emph{in
both phases of the cycle}, whereas the diffusion-dominated case allows correlations to
propagate only in the poloidal-to-toroidal phase. The other correlation is broken in
the diffusion-dominated regime because the advection is short-circuited by direct
diffusion, which transports flux downwards and equatorward -- where it is cancelled by
oppositely signed flux from the other hemisphere. This explains how the correlations
can survive for multiple cycles in the advection-dominated regime, but not in the
diffusion-dominated regime.

\section{Conclusion} \label{sec:conclusion}

Significant uncertainties remain in our understanding of the physics of the solar
dynamo mechanism, implying that prediction of future solar activity based on physical
models is a challenging task. Here we have demonstrated how a flux-transport dynamo
model behaves differently in advection and diffusion dominated regimes. Such
differences, amongst others, have previously led to conflicting predictions of the
amplitude of Cycle 24. \citet{dikpati2006} use an advection-dominated model to predict a
much stronger cycle than Cycle 23, whereas \citet{choudhuri2007} use a
diffusion-dominated model to predict a much weaker cycle. The latter prediction is
somewhat similar in spirit to the precursor methods \citep{schatten2005,
svalgaard2005}, which use the polar field at cycle minimum to predict the amplitude of
the following cycle. Owing to the lack of observations of conditions inside the
convection zone, opinions differ as to whether the real solar dynamo is weakly or
strongly diffusive (e.g. \citealp{dikpati2006b}; \citealp{jiang2007}).

We find that for low circulation speeds $v_0$ (in the diffusion-dominated regime), the
cycle amplitude is an increasing function of $v_0$, as in the observations of
\citet{hathaway2003}. However, the amplitude curve has a turnover point and is a
decreasing function of $v_0$ at higher $v_0$ (in the advection-dominated regime),
opposite to the observed correlation. When the diffusivity in the convection zone is
increased, the location of this turnover moves to a higher $v_0$. Our extensive
analysis shows that this turnover corresponds to the transition between the diffusion
and advection dominated regimes. In the diffusion-dominated regime, faster circulation
means less time for decay of the poloidal field, leading to a higher cycle amplitude,
whereas in the advection-dominated regime diffusive decay is less important and a
faster circulation means less time to induct toroidal field, thus generating a lower
cycle amplitude. If the observed statistics of the past $12$ cycles as reported by
\citet{hathaway2003} reflect a true underlying trend, then our results imply that the
solar dynamo is in fact working in a regime which is dominated by diffusive flux
transport in the main body of the SCZ (although the cycle period is still governed by
the slow meridional circulation counterflow at the base of the SCZ). This conclusion
supports the analysis of \citet{jiang2007}.

Through a correlation analysis in a stochastically forced version of our model, we have
also explored the persistence of memory in the solar cycle for both the diffusion-dominated and advection-dominated regimes. It is this memory mechanism which is
understood to lead to predictive capabilities in $\alpha$-$\Omega$ dynamo models with
spatially segregated source regions for the $\alpha$ and $\Omega$ effects. This
understanding is based on the finite time delay required for flux transport to
communicate between these different source regions. We find that the polar field of
cycle $n$ correlates strongest with the amplitude (toroidal flux) of cycle $n+1$ in
both the regimes. In the diffusion-dominated regime this is the only significant
correlation, indicative of a one-cycle memory only. However, in the advection-dominated
case, there are also significant correlations with the amplitude of cycles $n$, $n+2$,
and $n+3$. In contrast to the correlations that we infer, \citet{charbonneau2000} found
that the strongest correlation in their advection dominated model was with a two-cycle
time lag. Since such correlations lead to predictive capabilities, and obviously seem
to be model and parameter-dependent as suggested by our results, such a correlation
analysis should be the first step towards any prediction, the latter being based on the
former. In hindsight, however, both \citet{dikpati2006} -- who use an advection-dominated model and inputs from multiple previous cycles, and \citet{choudhuri2007} --
who use a diffusion-dominated model and input from only the past cycle to predict the
next cycle, appear to be have made the correct choices within their modelling
assumptions.

\begin{figure}
\plotone{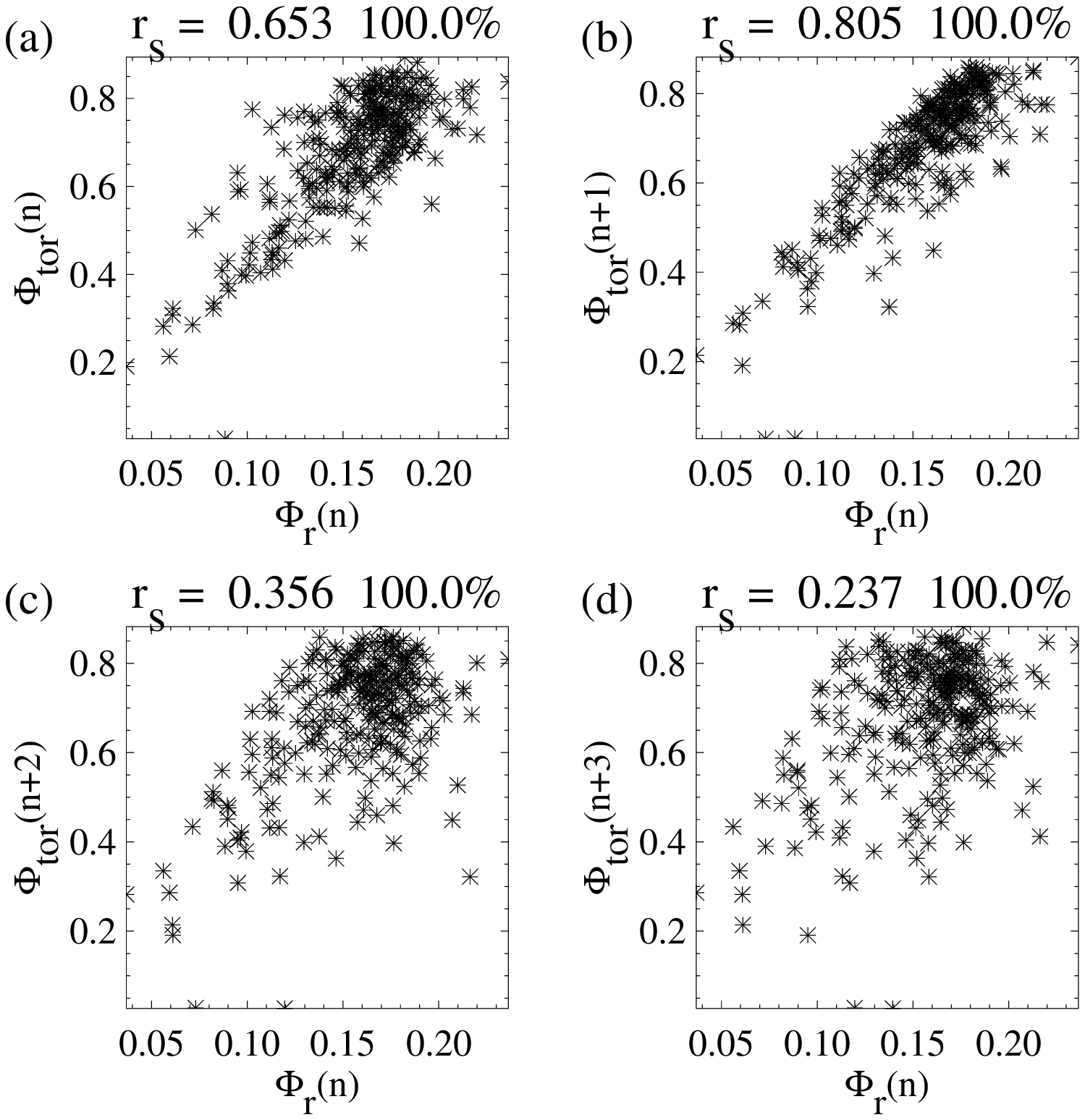}
\caption{Cycle-to-cycle correlations in the advection-dominated regime (run 2), between radial flux
$\Phi_r(n)$ and (a) toroidal flux $\Phi_{\textrm{tor}}(n)$, (b) $\Phi_{\textrm{tor}}(n+1)$,
(c) $\Phi_{\textrm{tor}}(n+2)$, and (d) $\Phi_{\textrm{tor}}(n+3)$. The Spearman's rank
correlation coefficient is given along with its significance level for 275 cycles. All
magnetic fluxes are in units of $10^{25}\mx$.} \label{fig:coradv}
\end{figure}

Note that the memory mechanism in our advection-dominated case appears to have a
different cause than that implied by \citet{dikpati2006b}, who invoke the survival of
multiple old-cycle polar fields feeding into a new cycle toroidal field. All of
the surviving correlations in our advection-dominated regime (Figure \ref{fig:coradv})
are positive; they do not alternate in sign. This alternation in sign would be expected if
bands of multiple older cycle poloidal field survive in the tachocline -- odd and even
cycle poloidal fields would obviously contribute oppositely because of their
alternating signs. In that case we would expect the absolute value of $\Phi_{r}(n)$ to
correlate positively with $\Phi_{\textrm{tor}}(n+1)$ and $\Phi_{\textrm{tor}}(n+3)$ and so
on, but negatively with $\Phi_{\textrm{tor}}(n+2)$ and $\Phi_{\textrm{tor}}(n+4)$ and so on,
as evident in the results of \citet[][Figure~9; after accounting for the fact that they use signed magnetic fields]{charbonneau2000}. Rather, in the advection-dominated regime
of our model, the correlations appear to persist simply because fluctuations in field
strength are passed on in both the poloidal-to-toroidal and toroidal-to-poloidal phases
of the cycle, as evidenced by the correlation between amplitude and polar flux of cycle
$n$. In a recent analysis, \citet{cameron2007} find that the predictive skill of a
surface flux transport model -- similar in spirit to the advection-dominated dynamo
model of \citet{dikpati2006} -- is contained in the input information of sunspot areas
in the declining phase of the cycle. They argue that memory of multiple past cycles, in
the form of surviving bands of poloidal field (its surface manifestations in their
case), need not be the only reason behind the predictive capability of the
advection-dominated dynamo model of \citet{dikpati2006}. Our analysis of the
advection-dominated regime supports this suggestion of \citet{cameron2007}.

Coming back to the diffusion-dominated regime, our comparative analysis indicates that
in this case, the memory of past cycles is governed by downward diffusion of poloidal
field into the tachocline -- which primarily results in a one-cycle memory. The fact
that diffusion is an efficient means for transporting flux is often ignored, especially
in this era of advection-dominated models; however, we find that diffusive flux
transport is quite efficient. The identification of this one cycle memory in our
stochastically forced model contradicts \citet{dikpati2006b} -- who claim that
prediction is not possible in this regime. As long as the source regions are spatially
segregated, and one of the source effects is observable and the other deterministic, flux
transport $\alpha$-$\Omega$ dynamos will inherently have predictive skills no matter
what physical process (i.e., circulation, or diffusion, or downward flux-pumping) is
invoked to couple the two regions. We may also point out that in the context of cycle-to-cycle correlations, downward flux pumping \citep{tobias2001} would have the same
effect as diffusion in that it also acts to short-circuit the meridional circulation
conveyor belt. So although downward flux pumping differs from diffusive transport
because in the latter case the fields may reduce in strength due to decay, the overall
persistence of memory is expected to be similar if diffusive flux transport was
replaced or complemented by downward flux pumping.

In summary, our analysis has served both to explore the diffusion dominated and
advection dominated regimes within the framework of a BL type dynamo, and to
demonstrate how the memory of the dynamo may be different in these two regimes. Based
on our analysis we assert that diffusive flux transport in the SCZ plays an important
role in flux transport dynamics, even if the dynamo cycle period is governed by the
meridional flow speed. In fact, the observed solar cycle amplitude-period dependence
may arise more naturally in the diffusion-dominated regime, as discussed earlier. Taken
together therefore, we may conclude that diffusive flux transport is a significant
physical process in the dynamo mechanism and this process leads primarily to a
one-cycle memory which may form the physical basis for solar cycle predictions, if
other physical mechanisms involved in the complete dynamo chain of events are well
understood. Separate, detailed examinations of these other related physical
mechanisms will be performed in the future.

\acknowledgements
This research was funded by NASA Living With a Star grant NNG05GE47G. We
also acknowledge support from the UK STFC and the Solar Physics NSF-REU program
at Montana State University.

\pagebreak


\clearpage
\begin{deluxetable}{lrrrrrrr}
\tablewidth{0pt}
\tablecaption{Cycle-to-cycle correlations \label{tab:correlations}}
\tablehead{ &
\multicolumn{3}{c}{$\Phi_r(n)$ for run 1} &&  \multicolumn{3}{c}{$\Phi_r(n)$ for run 2}  \\
                         & \multicolumn{3}{c}{(diffusion-dominated)} & \phantom{---} & \multicolumn{3}{c}{(advection-dominated)} \\
          & \multicolumn{2}{c}{$r_\textrm{s}$} & \multicolumn{1}{c}{$r_\textrm{p}$} && \multicolumn{2}{c}{$r_\textrm{s}$} & \multicolumn{1}{c}{$r_\textrm{p}$} }
\tablecolumns{8}
\startdata
$\Phi_{\textrm{tor}}(n)$   & 0.185 & $99.8\%$ \phantom{--}& \textit{0.287} &&  0.653 & $100.0\%$ \phantom{--}& \textit{0.778} \\
$\Phi_{\textrm{tor}}(n+1)$ & 0.737 & $100.0\%$ \phantom{--}& \textit{0.706} && 0.805 & $100.0\%$ \phantom{--}& \textit{0.851}\\
$\Phi_{\textrm{tor}}(n+2)$ & -0.040 & $49.1\%$ \phantom{--}& \textit{0.028} && 0.356 & $100.0\%$ \phantom{--}& \textit{0.546}\\
$\Phi_{\textrm{tor}}(n+3)$ & 0.195 & $99.9\%$ \phantom{--}& \textit{0.202} && 0.237 & $100.0\%$ \phantom{--}& \textit{0.417}\\
$\Phi_{\textrm{tor}}(n+4)$ & 0.036 & $44.5\%$ \phantom{--}& \textit{0.056} && 0.183 & $99.8\%$ \phantom{--}& \textit{0.357}\\
$\Phi_{\textrm{tor}}(n+5)$ & 0.107 & $92.3\%$ \phantom{--}& \textit{0.073} && 0.214 & $100.0\%$ \phantom{--}& \textit{0.316}\\
\enddata
\tablecomments{Correlation coefficients and significance levels for peak surface radial flux $\Phi_{r}$ versus peak toroidal flux $\Phi_{\textrm{tor}}$ for 275 cycles from stochastically forced dynamo simulations.}
\end{deluxetable}

\end{document}